\documentclass[preprint,3p,review]{elsarticle}
\graphicspath{ {./figures/} }
\usepackage{hyperref}
\usepackage{float}
\usepackage{verbatim} 
\usepackage{apalike}
\usepackage{amssymb}
\usepackage[cmex10]{amsmath}
\usepackage{mdwmath}
\usepackage{mdwtab}
\usepackage{eqparbox}
\usepackage{array}
\usepackage{mdwmath}
\usepackage{mdwtab}
\usepackage{eqparbox}
\usepackage{amsthm}

\usepackage{algpseudocode}
\usepackage{algorithm}
\usepackage{epstopdf}
\usepackage{multirow}
\usepackage{mathtools}
\usepackage{subcaption}
\usepackage{adjustbox}
\usepackage{longtable}
\graphicspath{ {./figures/} }
\usepackage{float}
\usepackage{verbatim} 
\usepackage{comment}
\usepackage{algpseudocode}
\usepackage{algorithm}
\usepackage{pdflscape}
\usepackage{url}
\usepackage{booktabs}
\usepackage{multirow}
\usepackage{lipsum}
\usepackage{calc,xcolor}
\usepackage{libertine}
\usepackage{booktabs} 
\usepackage{siunitx}  
\usepackage{caption}  
\usepackage[figuresright]{rotating}
\usepackage{bigints}
\usepackage{mathptmx}
\usepackage{graphicx}
\usepackage{epsfig}
\usepackage{amsmath}
\usepackage{euscript}
\usepackage{multirow}
\usepackage{makecell}
\usepackage{epstopdf}

\begin{document}
	
	\begin{frontmatter}
		%
		%
		
		\title{Linear stability analysis of non-isothermal plane Couette flow in an anisotropic and inhomogeneous porous layer underlying a fluid layer}
		
		\author{Nandita Barman}
		\author{Anjali Aleria}
		\author{Premananda Bera\corref{cor1}}
		\ead{pberafma@iitr.ac.in}
		
		\cortext[cor1]{Corresponding author.}
		\address{Department of Mathematics, Indian Institute of Technology Roorkee, Roorkee - 247667, India}
		\begin{abstract}
			This paper carries out a linear stability analysis of a plane Couette flow in a porous layer underlying a fluid layer where the porous layer is anisotropic and inhomogeneous. The plane Couette flow is induced due to the uniform movement of the upper plate and convection arises due to the maintenance of the temperature difference between the upper plate and the lower plate. The fluid considered is Newtonian and incompressible. Darcy model is used to narrate the flow in the porous layer and at the interface, the Beavers-Joseph condition is used. The Chebyshev collocation method is used to solve the generalized eigenvalue problem. Here, the effect of anisotropy and inhomogeneity of the porous medium, along with the ratio of the thickness of fluid to porous layer, i.e., depth ratio $(\hat{d})$, Reynolds number $(Re)$ and Darcy number $(\delta)$ are studied. The analysis is carried out majorly for water; however, the impact of anisotropy and inhomogeneity on different fluids by varying Prandtl number ($Pr$) is also studied. Depending on the value of parameters, the unimodal (porous mode or fluid mode), bimodal (porous mode and fluid mode) and also trimodal (porous mode, fluid mode and porous mode) nature of the neutral curve is obtained. The increasing value of the inhomogeneity parameter, depth ratio or decreasing value of the anisotropy parameter, Reynolds number, Prandtl number and Darcy number raises the system instability. For $\delta=0.002$, $Pr=6.9$ and $Re=10$, dominating nature of porous mode is always observed for $\hat{d}<0.07$, and fluid mode for $\hat{d}>0.21$ irrespective of anisotropy and inhomogeneity parameter. With the help of energy budget analysis, the types of instability are categorized and also the types of mode obtained from linear stability analysis are verified. Secondary flow patterns are also visualized to understand the flow dynamics. 
		\end{abstract}
	\end{frontmatter}
	
	\section{Introduction}\label{sec1}
	The transport phenomena in fluid overlying porous systems have gathered considerable interest over the last few decades due to their extensive applications in geophysics: the water flow under the Earth'surface and oil flow in underground reservoirs \cite{All-84, ALLEN1992125, Ewi-Wee-98} and industries: bioremediation of contaminated soil \cite{suchomel1998network}, construction of composite equipment in the automobile and aircraft industries \cite{Ble-Duf-Mck-Zul-99, Ble-Mck-Zul-Mar-99} etc. This motivates us to study this phenomenon under the application of non-isothermal Couette flow in such superposed systems. Further, the permeability of porous media,  in general, is anisotropic and inhomogeneous in nature, which hints to us to model the flow in an anisotropic and inhomogeneous porous layer \cite{nield2013convection}. Consequently, the present study deals with the hydrodynamic stability analysis of non-isothermal plane Couette flow in an anisotropic and inhomogeneous porous layer underlying the fluid layer. A brief account of literature in this direction is narrated below, which is described in a succeeding manner: first, fluid flow under natural convection; second: Poiseuille flow and Couette flow under non-isothermal conditions (mixed convection) and lastly, anisotropic and inhomogeneous porous media in both isothermal and non-isothermal cases.
	\par The first work on the onset of natural convection in a superposed fluid-porous layer was initiated by Sun \cite{Sun-73}. He reported a continuous decrement (increment) in the critical Rayleigh number in the porous (fluid) layer for the increasing depth ratio. Incorporating surface tension at the upper surface and considering constant heat-flux boundary conditions, Nield \cite{nield1977onset} obtained an analytical solution for the same. Chen and Chen \cite{Che-Che-88} investigated the combined impact of thermal and solutal diffusivity on the instability of the flow and found that the marginal stability curve exhibits bi-modality at low depth ratio $(\hat{d}\le 0.14)$. For $\hat{d}< 0.12$, the porous layer corresponding to the long-wave branch, whereas, for $\hat{d}>0.12$, the fluid layer corresponding to the short-wave branch controlled the system instability. The relative minimum in the long-wave region vanishes for large values of $\hat{d}$. Sun \cite{Sun-73} failed to obtain this result since he only observed the marginal stability curve in a limited range of wavenumber in the area of the long-wave critical point. Chen and Chen even validated their linear stability results experimentally \cite{Che-Che-89}. Further, the onset of convection in fluid overlying an anisotropic porous layer was studied by Chen \textit{et al.} \cite{chen1991convective} and they noticed that for the fixed value of the permeability of the porous medium along the vertical direction, decreasing the value of the ratio of horizontal to vertical permeability stabilizes the superimposed layer configuration by increasing its resistance to motion. In the porous layer, convection virtually disappears for $\hat{d} \ge 0.2$, resulting in motion primarily within the fluid layer. By considering inhomogeneous permeability in an anisotropic porous medium, Chen and Hsu \cite{Che-Hsu-91} extended the analysis of Chen \textit{et al.} \cite{chen1991convective}. They found that anisotropic and inhomogeneous effects are negligible when $\hat{d}\ge0.2$, and here, the convection is mostly limited to the fluid layer. Further, it was shown that the inhomogeneity causes convection to begin in the nearby region of higher permeability, which drives convection corresponding to a shorter critical wavelength. The most prominent findings from these literature, based on theoretical and experimental works, show that the depth ratio and media permeability comprising of anisotropy and inhomogeneity play a major role in determining the stability of the flow. 
	\par Application of shear flow with respect to pressure gradient (i.e., Poiseuille flow) and movement of the upper plate (i.e., Couette flow) in thermal convection is the topic of considerate discussion recently. As it is known that plane Poiseuille flow in the channel is linearly unstable for Reynolds number less than $5772$ and plane Couette flow in the channel is linearly stable for all values of Reynolds number \cite{orszag1980}, this fact itself shows that the study of these flows is independent and none of the results for one flow can be predicted from the results of other flow. Moreover, Chang investigated the thermal convection of plane Couette flow \cite{Cha-05} and Poiseuille flow \cite{Cha-06} in such a superposed fluid-porous system and the studies showed noteworthy differences between the Couette and Poiseuille flow, which are as follows: For decreasing value of $\hat{d}$, an increasing trend of oscillatory frequency as well as wave speed were observed in Couette flow \cite{Cha-05} whereas, the opposite characteristics were observed in Poiseuille flow \cite{Cha-06}. Additionally, while in Poiseuille flow \cite{Cha-06}, the oscillatory frequency variation for the fluid layer mode with respect to $Re$ is relatively small, as compared to the same in Couette flow \cite{Cha-05}. Further, Yin \textit{et al.} \cite{yin2013thermal} and Yin \textit{et al.} \cite{Yin-Wan-Wan-20} extended the studies \cite{Cha-05, Cha-06} by considering viscoelastic fluid and they observed the preferred mode of convection as the transverse mode in the presence of properly considered values of the parameter for viscoelastic fluid, in contrary to Newtonian fluid \cite{Cha-05, Cha-06}. 
	\par The porous media attributes, namely the anisotropy and inhomogeneity, have significantly affected the stability mechanism in isothermal flows in fluid overlying porous layers \cite{Dee-Ana-Bas-15}. Deepu \textit{et al.} observed that the system stability increases (decreases) for increasing the value of the anisotropy (inhomogeneity) parameter. Recently, Anjali \textit{et al.} \cite{anjalikhanbera2022} studied the stability analysis of plane Poiseuille flow in fluid overlying an anisotropic and inhomogeneous porous layer. They noticed that increasing inhomogeneity and decreasing anisotropy favored porous mode instability. Additionally, they identified the least stable mode as fluid for $\hat{d}> 0.16$ and porous for $\hat{d}< 0.05$. It is important to mention here that substantial differences between the work of Anjali \textit{et al.} \cite{anjalikhanbera2022} and the present study are observed, which are detailed in \S\ref{sec3}. 
	\par The above literature indicates that the impact of anisotropy and inhomogeneity under the shear generated by the movement of the upper plate and the inclusion of temperature difference between the upper and lower plate is not yet explored. It was also seen that the effect of media permeability was somehow not studied by Chang \cite{Cha-05}. The present study also directs on how it will affect the system's stability under both isotropic (anisotropic) and homogeneous (inhomogeneous) porous media. There is an interesting question of whether the experimental result of \cite{Che-Che-89} still holds under the present situation. The literature also poses the question of how these parameters impact the pattern of secondary flow and the mode of instability. Moreover, what causes the underlying type of instability? The present study aims to address the above questions.
	\par The article is put together as follows: The description of the physical problem and the mathematical formulation is given in \S\ref{sec2}, the results and discussions are presented in \S\ref{sec3}, whereas the concluding remarks are made in \S\ref{sec4}.   
	\section{Formulation of the problem}\label{sec2}
	\subsection{Physical Problem and Governing equations} 
\begin{figure}
		\centering
		\includegraphics[width=15cm, height=6.9cm]{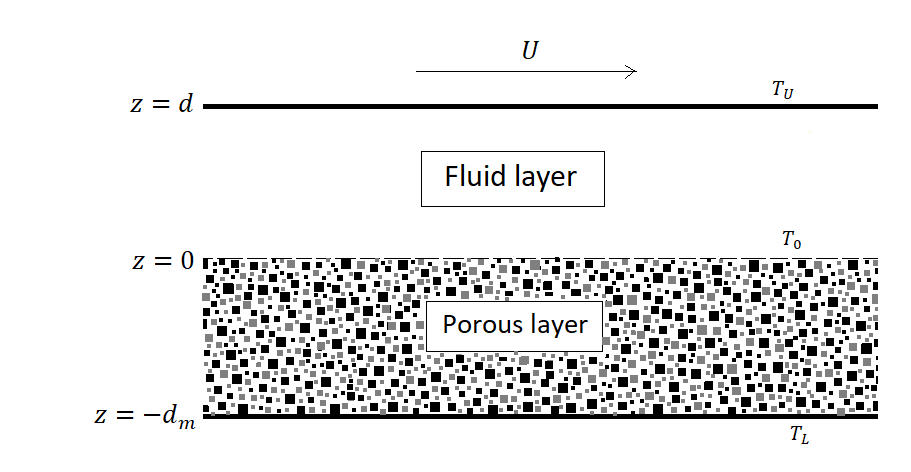}
		\caption{Diagram of the system}
		\label{fig:figure}
	\end{figure}

	The schematic of the problem under consideration is shown in Fig. \ref{fig:figure}. Here, a horizontal fluid layer of thickness $d$ overlying an anisotropic and inhomogeneous porous layer of thickness $d_m$ is considered where the interface between these two layers is located at $z=0$. The temperature of the upper plate is maintained at $T_U$, which is lower than the temperature of the lower plate $(T_L)$. $T_0$ is the reference temperature at $z=0$ so that $T_U<T_0<T_L$ and the upper plate of the fluid layer is moving with constant velocity $U$. The fluid considered here is viscous, Newtonian, incompressible and satisfies Boussinesq approximation. 
	The dimensional mass, momentum and energy balance equations for the fluid layer \cite{Cha-05} in the cartesian coordinate system are written as follows:
	\begin{align}
		&	\frac{\partial u}{\partial x}+\frac{\partial w}{\partial z}=0~, \label{eqn1} \\
		&	\frac{\partial u}{\partial t}+u\frac{\partial u}{\partial x}+w\frac{\partial u}{\partial z}=-\frac{1}{\rho_0}\frac{\partial p}{\partial x}+\nu \Delta u~, \label{eqn2} \\
		&	\frac{\partial w}{\partial t}+u\frac{\partial w}{\partial x}+w\frac{\partial w}{\partial z}=-\frac{1}{\rho_0}\frac{\partial p}{\partial z}+\nu \Delta w-g[1-\beta_T(T-T_0)]~, \label{eqn3} \\
		&	\frac{\partial T}{\partial t}+u\frac{\partial T}{\partial x}+w\frac{\partial T}{\partial z}=\alpha \Delta T~, \label{eqn4}
	\end{align}
	with time $t>0$ and $\{x\in\mathbb{R}, z\in[0,d]\}$. Here, the horizontal and vertical components of the velocity are denoted by $u$ and $w$ respectively, $T$ is the temperature, $p$ is the pressure for the fluid layer, $\nu$ denotes the kinematic viscosity of the fluid, $\rho_0$ is the density of the fluid at the reference temperature $T_0$, $\beta_T$ is the coefficient of thermal expansion, $\alpha$ is the thermal diffusivity, $g$ is the gravity, and $\Delta$ denotes the two-dimensional Laplacian operator.\\
	The value of porosity taken into account in the present analysis is $0.3$, which is small. So the equations of motion are narrated by  Darcy's law \cite{Cha-06, Hil-Stra-09, Anj2022, SAMANTA2022104230}. The dimensional mass, momentum and energy balance equations for the porous layer  \cite{anjalikhanbera2022} in the cartesian coordinate system are written as follows: 
	\begin{align}
		&	\frac{\partial u_m}{\partial x}+\frac{\partial w_m}{\partial z}=0~, \label{eqn5} \\
		&	\frac{1}{\chi}\frac{\partial u_m}{\partial t}=-\frac{1}{\rho_0}\frac{\partial p_m}{\partial x}-\frac{\nu}{K_x \eta_x(z/d_m)}u_m~, \label{eqn6}\\
		&	\frac{1}{\chi}\frac{\partial w_m}{\partial t}=-\frac{1}{\rho_0}\frac{\partial p_m}{\partial z}-\frac{\nu}{K_z \eta_z(z/d_m)}w_m-g[1-\beta_T(T_m-T_0)]~, \label{eqn7}\\
		&	G_m\frac{\partial T_m}{\partial t}+u_m\frac{\partial T_m}{\partial x}+w_m\frac{\partial T_m}{\partial z}=\alpha_m \Delta T_m~, \label{eqn8}
	\end{align}
	with $t>0$ and $\{x\in \mathbb{R}, z\in[-d_m, 0]\}$. The horizontal and vertical components of the seepage velocity are denoted by $u_m$ and $w_m$, respectively, in the porous medium. $\eta_x$ \& $\eta_z$ denote the inhomogeneity function and $K_x$ \& $K_z$ denote the permeability in $x$ and $z$ direction, respectively. The parameter $\chi$ denotes the porosity, $K^*=K_x/K_z$ is the anisotropy parameter, $\alpha_m=\kappa^*/\rho_0 c_p$ and $G_m=(\rho_0 c_p)^*/(\rho_0 c_p)$ with $ X^*=\chi X+(1-\chi)X_m$, where $X$ is substituted by $\kappa$ or $\rho_0 c_p$, $c_p$ is the specific heat of the fluid and $\kappa$ denotes the respective thermal conductivities.  \\
	The boundary conditions are as follows:\\
	At the upper plate of the fluid layer $z=d$, 
	\begin{equation}
		u=U,\quad w=0,\quad T=T_U~. \label{eqn9}
	\end{equation}
	At the lower plate of porous layer  $z=-d_m$,
	\begin{equation}
		w_m=0,\quad T_m=T_L~. \label{eqn10}
	\end{equation}
	At the interface of fluid and porous layers, i.e. at $z=0$, 
	\begin{equation}
		w=w_m,\quad T=T_m,\quad \alpha \frac{\partial T}{\partial z}=\alpha_m \frac{\partial T_m}{\partial z},\quad p-2\mu\frac{\partial w}{\partial z}=p_m~, \label{eqn11}  
	\end{equation} 
	\begin{equation}
		\frac{\partial u}{\partial z}+J\frac{\partial w}{\partial x}= \frac{\alpha_{BJ}}{\sqrt{K_x \eta_x(0)}}(u-u_m)~, \label{eqn12}
	\end{equation}
	where, equation \ref{eqn12} represents the Jones \cite{jones1973low} and Beavers-Joseph \cite{Bea-Jos-67} condition when $J=1$ and $J=0$, respectively and $\alpha_{BJ}$ denotes the Beavers-Joseph constant. Due to the lack of unified theory for fluid-porous interface conditions, several attempts in the literature have been made to derive the same \cite{jones1973low, Bea-Jos-67, Cha-Che-Str-06, Straughan_book, WU2019221}. Recently, \cite{mccurdy2019convection} have studied linear as well as non-linear stability analysis of non-isothermal flow in a coupled free fluid-porous system following \cite{girault2009dg}. At the interface between fluid and porous layers, Beaver-Joseph-Saffman condition was used and they came to the conclusion that the relative difference between the marginal stability curve obtained using the Beavers-Joseph, the Jones and the Beaver-Joseph-Saffman conditions is almost negligible for small Darcy number $(\delta=\frac{K_x^{1/2}}{d_m})$ i.e. for $\delta\le 5\times 10^{-3} $. Also, our numerical investigation reveals that the contribution of the term $\frac{\partial w}{\partial x}$ on the instability of the system is insignificant for the ranges of parameters taken into consideration, which is shown in \ref{sec6}. So, in the present study, the Beavers-Joseph condition $(J=0)$ is considered at the fluid-porous interface \cite{Cha-Che-Str-06}.
	\subsection{The basic flow}
	Assumption of the flow as unidirectional, steady and fully developed yields the basic analytical solution as follows: \\
	In the fluid layer, 
	\begin{equation}
		\overline{u}(z) =\left(\frac{\alpha_{BJ}U}{\sqrt{K_x\eta_x(0)}+\alpha_{BJ}d}\right)z+\frac{U\sqrt{K_x\eta_x(0)}}{\sqrt{K_x\eta_x(0)}+\alpha_{BJ}d},\quad \overline{w}=0~, \label{eqn13}
	\end{equation}
	\begin{equation}
		\overline{T}(z)=\left(\frac{T_U-T_0}{d}\right)z+T_0~. \label{eqn14}
	\end{equation}
	In the porous layer, 
	\begin{equation}
		\overline{u}_m=0,\quad \overline{w}_m=0~, \label{eqn15}
	\end{equation}
	\begin{equation}
		\overline{T}_m(z)=\left(\frac{T_0-T_L}{d_m}\right)z+T_0~. \label{eqn16}
	\end{equation}
	It is to be noted that, in contrast to isothermal/non-isothermal Poiseuille flow in a superposed fluid-porous system, where the basic flow in the porous layer is non-zero, here, for plane Couette flow, the same is zero. It gives a hint that the analysis found for Poiseuille flow \cite{anjalikhanbera2022} may differ significantly for Couette flow.
	\subsection{ Linear disturbance equation}
	The governing equations are nondimensionalized by $d,$ $U,$ $d/U,$ $(T_0-T_U)\nu/\alpha$ and $\mu U/d$ for fluid layer and $d_m,$ $\nu/d_m,$ $d_m^2/\nu,$ $(T_L-T_0)\nu/\alpha_m$ and $\rho_0(\nu/d_m)^2$ for the porous layer using the corresponding scales of length, velocity, time, temperature and pressure, respectively. To carry out the linear stability analysis, we introduce two-dimensional infinitesimal perturbations on the fully developed laminar base flow as given below
	\begin{equation}\label{eqn17}
		\begin{split}
			& 	(u,w,T,p)=(\overline{u},0,\overline{T},\overline{P})+(u',w',T',p'),\\ 
			&  	(u_m,w_m,T_m,p_m)=(0,0,\overline{T}_m,\overline{P}_m)+(u_m',w_m',T_m',p_m'),
		\end{split}
	\end{equation}
	where the prime quantities denote infinitesimal perturbation.\\
	Neglecting the nonlinear terms, dimensionless linearized perturbed equations in the fluid layer, $z\in[0,1]$, are :
	\begin{align}
		&	\frac{\partial{u'}}{\partial x}+\frac{\partial{w'}}{\partial z}=0~, \label{eqn18} \\
		&	Re \left(\frac{\partial{u'}}{\partial t}+\overline{u}\frac{\partial{u'}}{\partial x} +{w'}\frac{d\overline{u}}{dz}\right)=-\frac{\partial {p'}}{\partial x}+\Delta {u'}~, \label{eqn19}\\
		&	Re \left(\frac{\partial{w'}}{\partial t}+\overline{u}\frac{\partial{w'}}{\partial x}\right)=-\frac{\partial {p'}}{\partial z}+\Delta {w'}+\frac{Ra}{Re}{T'}~, \label{eqn20}\\
		&	Pr Re\left(\frac{\partial{T'}}{\partial t}+\overline{u}\frac{\partial{T'}}{\partial x}+ {w'}\frac{\partial \overline{T}}{\partial z}\right) = \Delta {T'}~, \label{eqn21}
	\end{align}
	where
	\begin{equation} \label{eqn22}
		Re=\frac{dU}{\nu},\quad Ra=\frac{g\beta_T(T_0-T_U)d^3}{\nu \alpha},\quad Pr=\frac{\nu}{\alpha}.
	\end{equation}
	In the above Eq. \ref{eqn22}, $Re$, $Ra$, and $Pr$ denote the Reynolds number, Rayleigh number and Prandtl number in the fluid layer, respectively.\\
	In the porous layer, $z_m\in[-1,0]$, the linearized perturbed equations are
	\begin{align}
		&	\frac{\partial{u}_m'}{\partial x_m}+\frac{\partial{w}_m'}{\partial z_m}=0~, \label{eqn23}\\ 
		&	\frac{1}{\chi}\frac{\partial{u}_m'}{\partial t_m}=-\frac{\partial{p}_m'}{\partial x_m}-\frac{1}{\delta^2}\frac{{u}_m'}{\eta_x}~, \label{eqn24}\\
		&	\frac{1}{\chi}\frac{\partial{w}_m'}{\partial t_m}=-\frac{\partial {p}_m'}{\partial z_m}-\frac{K^*}{\delta^2}\frac{{w}_m'}{\eta_z}+\frac{Ra_{m}}{\delta^2}{T}_m'~, \label{eqn25}\\
		&   Pr_m \left(G_{m}\frac{\partial {T}_m'}{\partial t_m}+{w}_m'\frac{\partial \overline{T}_m}{\partial z_m}\right)= \Delta^m {T}_m' ~, \label{eqn26}
	\end{align}
	where 
	\begin{equation}\label{eqn27}
		\delta=\frac{K_x^{1/2}}{d_m},\quad Pr_m=\frac{\nu}{\alpha_m},\quad Ra_m=\frac{g\beta_T(T_L-T_0)d_m K_x}{\nu \alpha_m}.
	\end{equation}
	In the above Eq. \ref{eqn27}, $\delta$, $Pr_m$ and $Ra_m$ denote the Darcy number, Prandtl number and Rayleigh number in the porous layer, respectively. The relation between the parameters $Pr$, $Pr_m$ and $Ra$, $Ra_m$ is as follows:
	\begin{equation}\label{eqn28}
		Pr_m=\epsilon Pr,\quad Ra_m=\dfrac{\epsilon^2\delta^2}{\hat{d}^4}Ra,
	\end{equation}
	where $\epsilon=\alpha/\alpha_m$ is the ratio of thermal diffusivity and $\hat{d}=d/d_m$ is the depth ratio.
	
	The dimensionless boundary conditions on the upper plate, $z=1$, are 
	\begin{equation}\label{eqn29}
		u'=w'=T'=0,
	\end{equation}
	and on the bottom plate, $z_m=-1$,
	\begin{equation}\label{eqn30}
		w_m'=T_m'=0,
	\end{equation}
	On the interface, $z=z_m=0$,
	\begin{align}\label{eqn31}
		Rew'=\hat{d} w_m', \hspace{0.5cm} \hat{d} T'= \epsilon^2 T_m',  \hspace{0.5cm} \frac{\partial T'}{\partial z}
		= \epsilon \frac{\partial T_m'}{\partial z_m},
   \end{align}
  \begin{align}\label{eqna}
		\frac{\partial u'}{\partial z}=\frac{\alpha_{BJ} \hat{d}}{\delta \sqrt{\eta_x(0)}}\left(u'-\frac{\hat{d}^2}{Re}u_m'\right),  \hspace{0.5cm} p'= 2 \frac{\partial w'}{\partial z} + \frac{\hat{d}^2}{Re} p_m'.
	\end{align}
	Using normal mode analysis \cite{Dra-Rei-04}, the disturbances can be written as 
	\begin{align}	
		&  (w', T', p')=[W(z), \theta(z), \pi(z)]\exp(-i\sigma t + i ax),\label{eqn32}\\
		&  ({w}_m', {T}_m', p_m')=[{W}_m(z_m), \theta_m(z_m), \pi(z_m)]\exp(-i\sigma_m t_m + i a_mx_m).\label{eqn33}
	\end{align}
	Here $a$ and $a_m$ are the real-valued wave number in the horizontal direction for the fluid and porous layer, respectively, and the complex wave speed for the fluid and porous layer is $\sigma=\sigma^r+i\sigma^i$ and $\sigma_m=\sigma_m^r+i\sigma_m^i$, respectively. Substituting Eqs. \eqref{eqn32}-\eqref{eqn33} into \eqref{eqn18}-\eqref{eqn21}, \eqref{eqn23}-\eqref{eqn26}, \eqref{eqn29}-\eqref{eqna} and eliminating pressure terms, the linearized disturbance equations in the fluid layer are
	\begin{align}
		&	(D^2-a^2)^2 W-i a Re \overline{u}(D^2-a^2)W -\frac{Ra}{Re} a^2 \theta=-i\sigma Re (D^2-a^2)W~,\label{eqn34}\\
		&	(D^2-a^2) \theta-i a Re Pr \overline{u}\theta+Re W =-i\sigma Pr Re \theta~,\label{eqn35}
	\end{align}
	and in the porous layer are
	\begin{align}
		&	\dfrac{D_m^2 W_m}{\eta_x}-\dfrac{D_m W_m}{\eta_x^2}D_m\eta_x-\dfrac{K^* a_m^2}{\eta_z}W_m+Ra_m a_m^2\theta_m=i\dfrac{\sigma_m \delta^2}{\chi}(D_m^2-a_m^2)W_m~,\label{eqn36}\\
		&	(D_m^2-a_m^2) \theta_m + W_m  = -i\sigma_m G_m Pr_m \theta_m~.\label{eqn37}
	\end{align}
	
	The boundary conditions at $z=1$ are
	\begin{equation}\label{eqn38}
		W=DW=\theta=0~,
	\end{equation}
	and at $z_m=-1,$ 
	\begin{equation}\label{eqn39}
		W_m=\theta_m=0.
	\end{equation}
	At the interface, $ z=z_m=0 $,
	\begin{align}
		& Re W=\hat{d}W_m, \quad \hat{d}\theta=\epsilon^2\theta_m, \quad D\theta=\epsilon D_m\theta_m,\label{eqn40}\\
		& D^2 W = \frac{\alpha_{BJ}\hat{d}}{\delta \sqrt{\eta_x(0)}}\left(DW-\frac{\hat{d}^2}{Re}D_m W_m\right),\label{eqn41}\\
		&	D^3W-3a^2DW-i a Re \overline{u}DW+i aRe\frac{d\overline{u}}{dz}W+\frac{\hat{d}^4}{Re\delta^2\eta_x}D_m W_m=-i\sigma Re DW+i\sigma_m\frac{\hat{d}^4}{\chi Re}D_m W_m.	\label{eqn42}
	\end{align}
	Here,
	\begin{equation}\label{eqn43}
		D=\dfrac{d}{dz},\quad D_m=\dfrac{d}{dz_m},\quad a=\hat{d}a_m,\quad \sigma=\dfrac{\hat{d}^2}{Re}\sigma_m.
	\end{equation}
	\par Applying the Chebyshev spectral collocation method with Chebyshev polynomials as the basis function, the aforementioned system of linear differential Eqs. \eqref{eqn34}-\eqref{eqn37} and boundary conditions \eqref{eqn38}-\eqref{eqn42} are discretized at the Gauss-Lobatto points \cite{Canuto-88}. By considering $\zeta=2z-1$ and $\zeta_m=-2z_m-1$, respectively, the fluid domain, $[0,1]$, and the porous domain, $[-1,0]$, are transformed \cite{Kha-Ber-20} to the Chebyshev domain $[-1,1]$.
	
	\par The discretized equations are written as a generalized eigenvalue problem of the form
	\begin{equation}\label{eqn44}
		\mathcal{A}\mathbb{X}=\sigma\mathcal{B}\mathbb{X},
	\end{equation} 
	where $\sigma$ and $\mathbb{X}$ denote the eigenvalue and eigenvector of field entities, respectively, and  $\mathcal{A}$ and $\mathcal{B}$ are the complex square matrices. MATLAB's built-in QZ algorithm \cite{Mol-Ste-73} is used to determine the generalized eigenvalues.
	\subsection{Kinetic Energy Spectrum}
	\par The physical mechanism behind the instability induced can be described by the energy budget analysis. As proposed by  \cite{Hop-Boy-83, Boo-Mie-96, SAHU2011987, Sha-Kha-Ber-18}, the rate of change of disturbance kinetic energy for the fluid layer is given by 	
	\begin{eqnarray}\label{eqn45}
		\dfrac{1}{2 \lambda}\bigintss_{0}^{1} \bigintss_{0}^{\lambda}\dfrac{\partial}{\partial t} \left(u'^2+w'^2\right) \,dx \,dz =-\dfrac{1}{\lambda}\bigintss_{0}^{1} \bigintss_{0}^{\lambda} u'w' \left(\dfrac{d\bar{u}}{dz}\right) \,dx \,dz-\dfrac{1}{\lambda Re}\bigintss_{0}^{1} \bigintss_{0}^{\lambda} \bigg[\left(\dfrac{\partial u'}{\partial x}\right)^2\nonumber\\+ \left(\dfrac{\partial u'}{\partial z}\right)^2+\left(\dfrac{\partial w'}{\partial x}\right)^2+ \left(\dfrac{\partial w'}{\partial z}\right)^2\bigg] \,dx \,dz + \dfrac{Ra}{\lambda Re^2}\bigintss_{0}^{1} \bigintss_{0}^{\lambda} T'w' \,dx \,dz+\dfrac{1}{\lambda Re}\bigintss_{0}^{\lambda}\bigg[w'p'\bigg]_{z=0}\,dx\nonumber\\-\dfrac{1}{\lambda Re} \bigintss_{0}^{\lambda} \bigg[u'\dfrac{\partial u'}{\partial z} +w'\dfrac{\partial w'}{\partial z} \bigg]_{z=0} \,dx \,dz,\nonumber\\
	\end{eqnarray}
	where, $\lambda=2\pi/a.$\\
	For the porous layer,
	\begin{eqnarray}\label{eqn46}
		\dfrac{1}{2 \lambda_m \chi}\bigintss_{-1}^{0} \bigintss_{0}^{\lambda_m} \dfrac{\partial}{\partial t_m} \left({u_m}'^2+w_m'^2\right)\,dx_m \,dz_m =\dfrac{ Ra_m}{\lambda_m\delta^2 }\bigintss_{-1}^{0} \bigintss_{0}^{\lambda_m}{ T_m' w_m'}\,dx_m \,dz_m-\nonumber\\\dfrac{1}{\lambda_m\delta^2}\bigintss_{-1}^{0} \bigintss_{0}^{\lambda_m}\bigg[\dfrac{u_m'^2}{ \eta_x}+\dfrac{K^* w_m'^2}{\eta_z} \bigg] \,dx_m \,dz_m-\dfrac{1}{\lambda_m }\bigintss_{0}^{\lambda_m}\bigg[w_m'p_m'\bigg]_{z_m=0}\,dx_m,\nonumber\\
	\end{eqnarray}
	where, $\lambda_m=2\pi/a_m$.
	\par Combining Eqs.  \eqref{eqn45} and  \eqref{eqn46}, the energy budget equation takes the standard form \cite{Ber-Kha-02,Kha-Ber-Kha-19}
	\begin{equation}\label{eqn47}
		\dfrac{\partial\left(KE\right)}{\partial t}+\dfrac{\partial(KE_m)}{\partial t_m}=E_s+E_b+E_d+I+E_{bm}+E_{Dm},
	\end{equation}
	where,
	\begin{equation}\label{eqn48}
		KE=\dfrac{1}{2 \lambda}\bigintss_{0}^{1} \bigintss_{0}^{\lambda} \left(u'^2+w'^2\right) \,dx \,dz , 
	\end{equation}
	and
	\begin{equation} \label{eqn49}
		KE_m=\dfrac{\hat{d}^3}{2 \lambda_mRe^3 \chi}\bigintss_{-1}^{0} \bigintss_{0}^{\lambda_m}\left(u_m'^2+w_m'^2\right) \,dx_m \,dz_m ,
	\end{equation}
	represent the disturbance kinetic energies for the fluid and porous layer, respectively, and
	\begin{equation}\label{eqn50}  
		E_s= -\dfrac{1}{\lambda}\bigintss_{0}^{1} \bigintss_{0}^{\lambda} u'w' \left(\dfrac{d\bar{u}}{dz}\right) \,dx \,dz,
	\end{equation}
	represents the amount of energy transfer between base state and the disturbed state via the Reynolds stress,
	\begin{equation} \label{eqn51} 
		E_b=\dfrac{Ra}{\lambda Re^2}\bigintss_{0}^{1} \bigintss_{0}^{\lambda} T'w' \,dx \,dz ,
	\end{equation}
	\begin{equation} \label{eqn52}
		E_{bm}=\dfrac{\hat{d}^3 Ra_m}{\lambda_mRe^3\delta^2 }\bigintss_{-1}^{0} \bigintss_{0}^{\lambda_m}{ T_m' w_m'}\,dx_m \,dz_m,
	\end{equation}
	are the disturbance kinetic energy terms due to work done by buoyancy for fluid and porous layers, respectively,
	\begin{align} \label{eqn53}
		E_d= -\dfrac{1}{\lambda Re}\bigintss_{0}^{1} \bigintss_{0}^{\lambda} \bigg[\left(\dfrac{\partial u'}{\partial x}\right)^2+ \left(\dfrac{\partial u'}{\partial z}\right)^2+\left(\dfrac{\partial w'}{\partial x}\right)^2+ \left(\dfrac{\partial w'}{\partial z}\right)^2\bigg] \,dx \,dz ,
	\end{align}
	represents the viscous dissipation in the fluid layer,
	
	\begin{align}  \label{eqn54}
		E_{Dm}=-\dfrac{\hat{d}^3}{\lambda_mRe^3\delta^2}\bigintss_{-1}^{0} \bigintss_{0}^{\lambda_m}\bigg[\dfrac{u_m'^2}{ \eta_x}+\dfrac{K^* w_m'^2}{\eta_z} \bigg] \,dx_m \,dz_m,
	\end{align} 
	represents the energy loss against surface drag and  
	\begin{align} \label{eqn55}
		I= \dfrac{1}{\lambda Re}\bigintss_{0}^{1}\bigg[w'p'- \left(u'\dfrac{\partial u'}{\partial z} +w'\dfrac{\partial w'}{\partial z}\right) \bigg]_{z=0} \,dx -\dfrac{\hat{d}^3}{\lambda_mRe^3 }\bigintss_{-1}^{0}\bigg[w_m'p_m'\bigg]_{z_m=0}\,dx_m,
	\end{align}
	denotes the work done due to the continuity of stresses and velocities at the interface of fluid and porous layers.  
	\section{Results and discussion}\label{sec3}
	A rigorous stability analysis has been made to understand the instability boundary of non-isothermal plane Couette flow in an anisotropic and inhomogeneous porous layer underlying a fluid layer. Note that the distribution of particle size in porous medium follows an exponential distribution  \cite{perkins1963review, doomra2022effect}. Consequently, the inhomogeneity function in $x$ and $z$ directions are taken as $\eta_x=e^{A(1+z_m)}$ and $\eta_z=e^{B(1+z_m)}$, respectively. It is observed that the effect of variation of $A$ by fixing $B$ and the effect of variation of $B$ by fixing $A$ are almost similar for the considered problem (see Fig.\ref{fig17} in \ref{sec5}). So, to reduce complications in regard to the directional inhomogeneities, $\eta_x$ and $\eta_z$ are taken to be identical in flow-normal and flow-parallel directions, i.e., $A=B$ is considered. At $z_m=-1$, $\eta_x=\eta_z=1$ and for positive (negative) values of $A$, the inhomogeneity in horizontal and vertical directions, i.e. $\eta_x$ and $\eta_z$ increase (decrease) vertically. Here, the flow is majorly governed by seven controlling parameters. The Darcy number $(\delta)$, permeability ratio or anisotropy parameter $(K^*)$ and inhomogeneity parameter (in terms of $A$) control the permeability of the porous layer whereas depth ratio $(\hat{d})$ characterizes the location of the interface, Reynolds number $(Re)$ determines the strength of the flow/forced flow, Prandtl number $(Pr/Pr_m)$ signifies the fluid type and Rayleigh number $(Ra/ Ra_m)$ determines the heat source intensity. Thus, the following study has been made for a good range of these parameters. Other parameters generally used in the literature \cite{Sun-73, Che-Che-88, str-01, Str-02, Cha-05, Cha-06} like porosity $(\chi)$, ratio of thermal diffusivity $(\epsilon)$, Beavers-Joseph constant $(\alpha_{BJ})$ and heat capacity ratio $(G_m)$ are fixed at $0.3,$ $0.7,$ $0.1$ and $10$, respectively. 
	\par The theoretical validation is done in the limiting case of isotropic and homogeneous porous medium with the results of \cite{Cha-05}, which perfectly matches with the published results and are given in Table \ref{theory}. At the same time, the experimental results are validated with the theoretical results of the present study for natural convection in isotropic and homogeneous porous media in the limit $Re\to 0$ by reproducing the results of \cite{Che-Che-89} and are given in Table \ref{experiment}. On the basis of various numerical experiments, it is found that 50 terms of the Chebyshev polynomial are enough to yield satisfactory convergence for the present study, so we have fixed the order of the polynomial at $50$. \par Before discussing the stability of the considered problem, a note is made regarding the possible impact of inhomogeneity and media anisotropy on the stability of isothermal plane Couette flow in the fluid overlying porous system. \cite{Cha-Che-Cha-17} observed that the isothermal plane Couette flow in the fluid overlying isotropic and homogeneous porous layer is unconditionally stable. Whether this observation will remain valid on introducing anisotropy and inhomogeneity in the porous medium? We have addressed this question and our numerical experiments in this direction reveal that the Couette flow in a fluid-porous system is also always linearly stable for considered values of anisotropy and inhomogeneity under isothermal conditions. 
	
\begin{table}[h]
	
	\caption{The values of critical $Ra_m, a_m$ and ${\sigma_m}^r$ obtained by Chang \cite{Cha-05} and present study for several $\hat{d}$ and $Re=10$, $Pr=6.5$, $\delta=0.002$, $\epsilon=0.7$, $\alpha_{BJ}=0.1$, $G_m=10$ and $\chi=0.3$.}  	
		\centering
		\begin{tabular}{lllllll}			
			\hline
			$\hat{d}$ & $Ra_m$ & $Ra_m$  & $a_m$ & $a_m$ & $-{\sigma_m}^r$ & $-{\sigma_m}^r$  \\ 
			& Chang\cite{Cha-05} & (present study) & Chang\cite{Cha-05}  & (present study)  &  Chang\cite{Cha-05} & (present study) \\ 
			\hline			      
			0.10 & 23.11 & 23.109660 & 2.12 & 2.12 & $-0.03627$ & $-0.036267$ \\
			0.12 & 22.55 & 22.545865 & 2.09 & 2.09 & $-0.04079$ & $-0.040793$\\
			0.14 & 13.31 & 13.312916 & 18.42 & 18.42 & $-704.832$ & $-704.832171$ \\
			0.16 & 8.04 &  8.035135 & 16.22 & 16.22 & $-537.339$ & $-537.338916$\\
			0.18 & 5.14 & 5.138917 & 14.49& 14.49 & $-423.208$ & $-423.207972$ \\
			0.20 & 3.44 &  3.440960 & 13.10 &  13.10 & $-342.127$ & $-342.126487$\\		
			\hline
		\end{tabular}
	\label{theory} 
\end{table}

\begin{table}[h]
	\caption{The values of critical $Ra_m$ obtained  experimentally by Chen \& Chen \cite{Che-Che-89} and present study for several $\hat{d}$. The other parametric values can be referred from Table $1$ and Table $2$ in Chen \& Chen \cite{Che-Che-89}}   
	
	\centering
	\begin{tabular}{lll}
		\hline
		$\hat{d}$ & $Ra_m$  & $Ra_m$\\ 
		& (present study)  &  Chen \& Chen \cite{Che-Che-89} \\ 
		\hline				
		0 & 39.477879  &  40.07\\
		0.025 & 28.145687  & 31.82\\ 
		0.1 & 19.152601  & 17.57\\
		0.2 & 3.521270  & 3.95\\
		0.5 & 0.158785  & 0.159\\
		1 & 0.012084  & 0.0124\\
		\hline
	\end{tabular}
\label{experiment}
\end{table}

\subsection{Effect of variation in anisotropy}
\begin{figure}
\centering
\includegraphics[width=\textwidth]{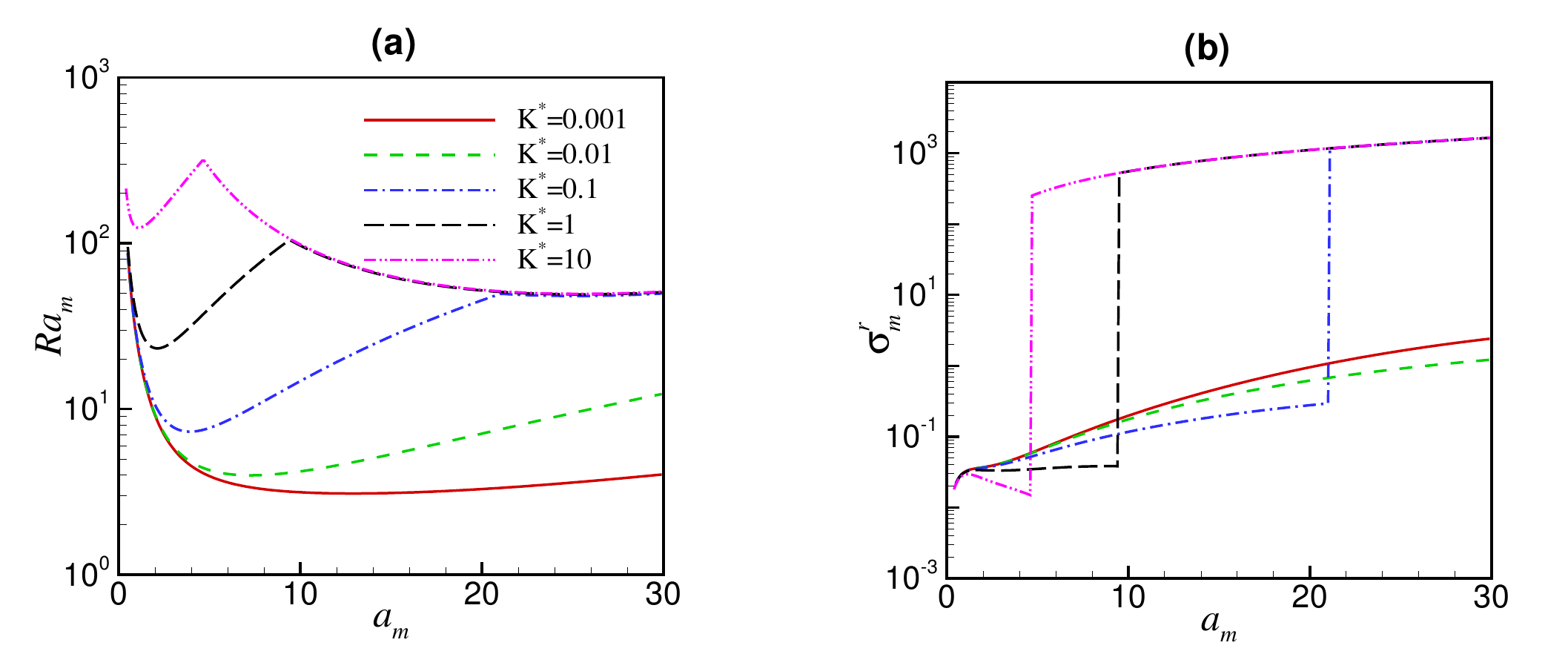}
\caption{Variation of neutral curves with corresponding oscillatory frequency for different anisotropy parameters at $\hat{d}=0.1$, $\delta=0.002$, $Pr=6.9$ and $Re=10$.}
\label{fig2}
\end{figure}
\begin{figure}
\centering
\includegraphics[width=\textwidth]{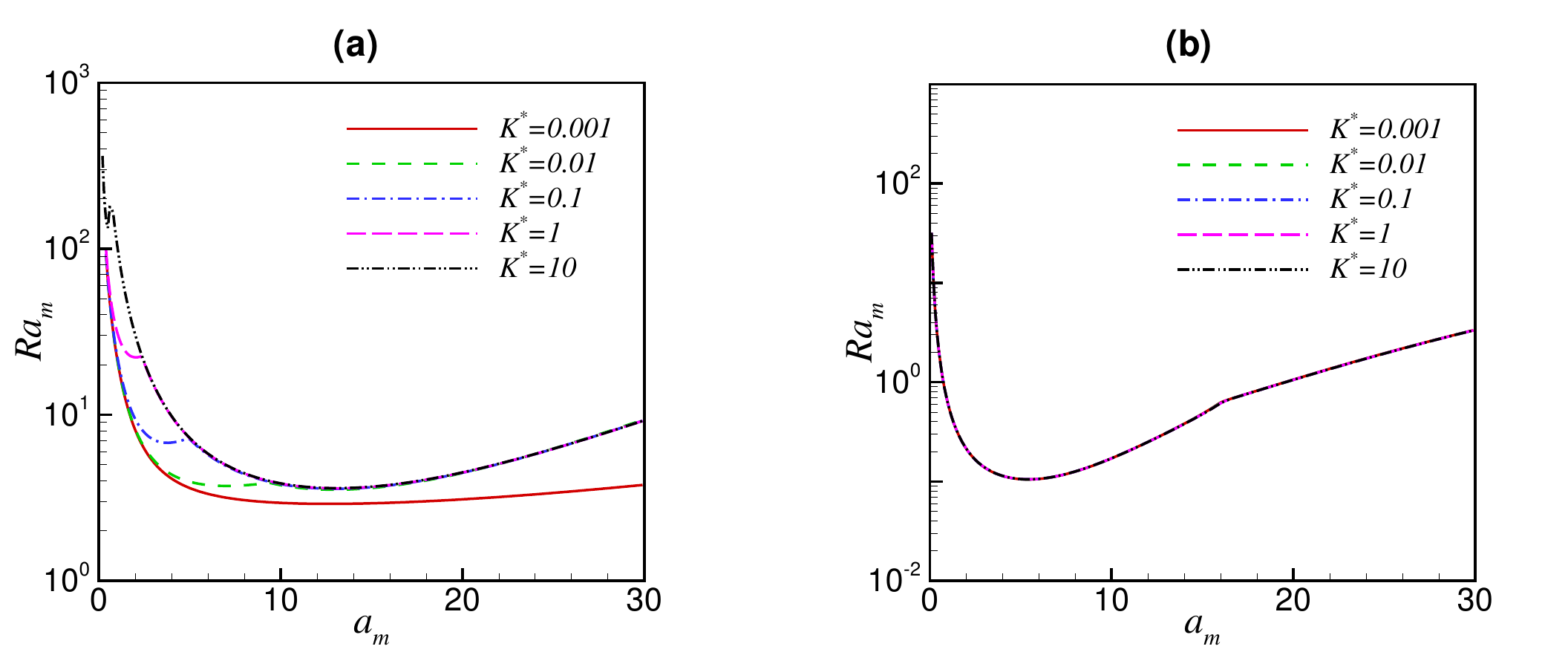}
\caption{Variation of neutral curves for different anisotropy parameter for (a) $\hat{d}=0.2$, (b) $\hat{d}=0.5$ with $\delta=0.002$, $Pr=6.9$ and $Re=10$.}
\label{fig3}
\end{figure}
To inspect the anisotropy effect on the instability behavior, we have considered porous medium as homogeneous, i.e., $\eta_x=\eta_z=1$ and have taken five values $0.001$, $0.01$, $0.1$, $1$ and $10$ of anisotropy parameter \cite{Dee-Ana-Bas-15}. Figure \ref{fig2} represents the variation of neutral curves with corresponding oscillatory frequency for several values of anisotropy parameter, $K^*$ at $\hat{d}=0.1$, $\delta=0.002$ and $Re=10$ for water. Different values of $K^*$ result in both unimodal and bimodal behavior of the neutral curve. The mode correlated with the lobe of the neutral curve at a small value of oscillatory frequency is named as the porous mode, and the mode correlated with the lobe of the neutral curve at a large value of oscillatory frequency is named as the fluid mode, where the convection is mostly confined to the porous layer and the fluid layer, respectively. It is noted that the neutral curve is unimodal for small values of $K^*$, i.e., for $0.001$ and $0.01$ and the porous mode entirely dominates the system instability. For the increasing value of $K^*$, the unimodal behavior of the neutral stability curve changes to bimodal. At $K^*=0.1$ and $1$, the porous mode is still dominating where the range of wave number corresponding to porous mode is $0\le a_m< 21.10$ and  $0\le a_m< 9.50$, respectively. But at $K^*=10$, the critical mode of convection changes to fluid mode, and the range of wave number corresponding to that mode is $4.70\le a_m\le 30$. Here, it is clear that the porous mode becomes more unstable for decreasing value of $K^*$, which is also true logically since a decrement in anisotropy causes the permeability in the vertical direction to increase (as $K_x$ is fixed), and as a result, flow resistance is reduced and consequently, the flow is destabilized in the porous layer. So, the introduction of anisotropy in the porous layer leads to earlier onset of convection than in the isotropic case obtained by \cite{Cha-05}. For example, the critical Rayleigh number obtained by \cite{Cha-05} was $23.11$ for $\hat{d}=0.1$, $\delta=0.002$, $Pr=6.5$ and $Re=10$. Whereas, after introducing anisotropy to the porous medium and maintaining the other parametric value the same, the critical Rayleigh number for $K^*=0.001$ obtained in the current study is $3.09$. 

\begin{figure}
\centering
\includegraphics[width=\textwidth]{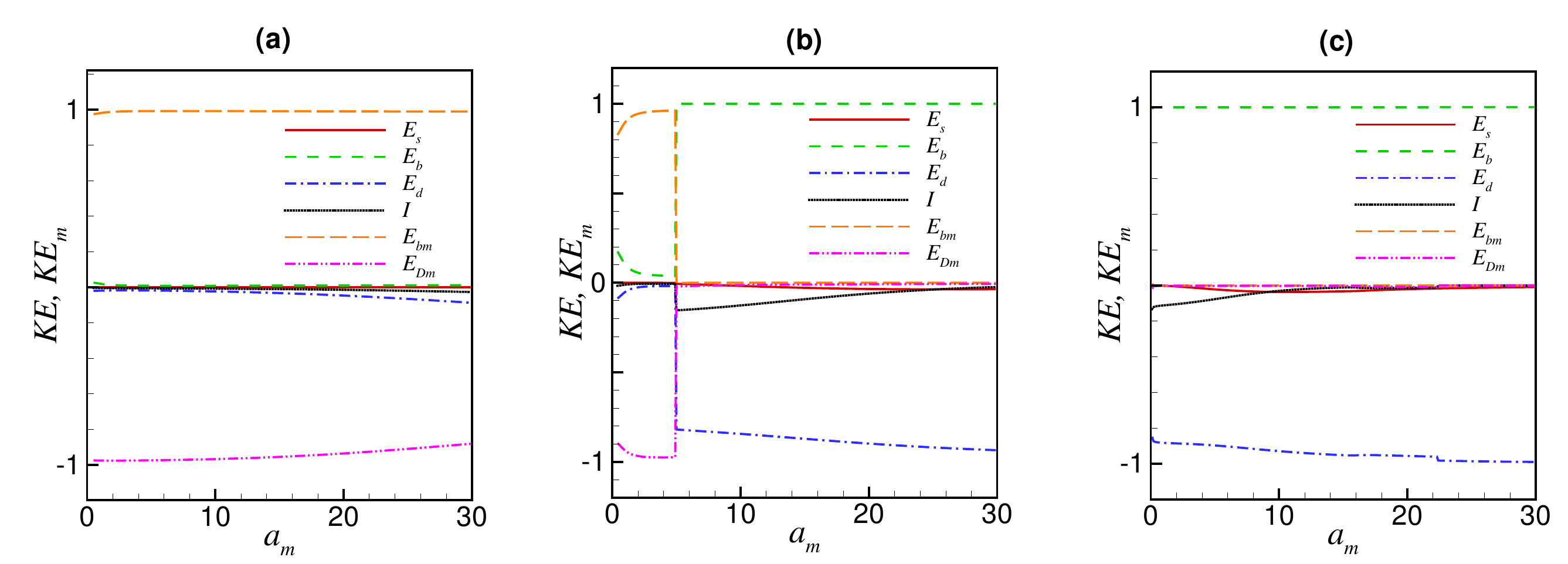}
\caption{Energy components (a) $\hat{d}=0.1$, $K^*=0.001$, (b) $\hat{d}=0.1$, $K^*=0.1$, (c) $\hat{d}=0.5$,$K^*=0.001$ with $\delta=0.002$, $Pr=6.9$ and $Re=10$.}
\label{fig4}
\end{figure}

As $\hat{d}$ is increased to $0.2$  (see Fig. \ref{fig3}(a)), the modal behavior of the neutral curve is consistent with Fig. \ref{fig2} for different values of $K^*$ except for $K^*=0.01$. For this value of $K^*$, the bimodal character of the stability curve is observed as contrary to the case for $\hat{d}=0.1$, where we have witnessed the unimodal character of the neutral curve. Also, we observe that fluid mode always dominates the instability of the system for $K^*>0.001$ at $\hat{d}=0.2$. So, a natural query comes to mind whether a further enlargement of $\hat{d}$, the unimodal characteristic of the neutral curve for $K^*=0.001$, will be changed to bimodal or not. To answer this, Fig. \ref{fig3}(b) is observed. At $\hat{d}=0.5$ (Fig. \ref{fig3}(b)), it is clear that there is no role of porous mode on the instability behavior of the system. Here, irrespective of the value of the anisotropy parameter $K^*$, the neutral curve only exhibits unimodal behavior, with fluid mode dominating the instability of the system and the role of this parameter on the instability of Couette flow dies out. 

To understand the underlying physical mechanism behind the instability behavior, we plot the kinetic energy (KE) balance spectrum, which will enable us to categorize the type of instability and verify the mode type obtained from the neutral stability curves. From our analysis of the KE balance spectrum, we have observed that there are only three positive terms that act as a destabilizing factor: $E_s$, $E_b$ \& $E_{bm}$ and on the basis of the contribution of these three terms, the type of instability is defined \cite{Kha-Ber-20b}. An instability is thermal-shear (thermal-buoyant in the fluid layer or thermal-buoyant in the porous layer) if $E_s$ ( $E_b$ or $E_{bm}$) contributes more than $70\%$ in the KE balance spectrum. To understand the underlying instability, we have considered three sets of values $(0.1, 10, 6.9, 0.002, 0.001)$, $(0.1, 10, 6.9, 0.002, 0.1)$, and $(0.5, 10, 6.9, 0.002, 10)$ of the parameters $(\hat{d}, Re, Pr, \delta, K^*)$.  Figure \ref{fig4} shows that the buoyancy effect remains positive in fluid and porous domains throughout, conclusively making buoyancy a destabilizing factor. Figure \ref{fig4}(a) shows that $E_{bm}$ is the most dominant term in the KE balance; thus, buoyancy affects the prevailing instability in the porous layer. In Fig. \ref{fig4}(b), the term $E_{bm}$ is most dominant for the range $0\le a_m<21.10$, and outside that range, the most dominant term is $E_b$. Thus thermal-buoyant instability in the porous layer is observed when  $0\le a_m<21.10$ and outside this range of wave number, thermal-buoyant instability in the fluid layer is observed, which was also obtained from the neutral stability curves (see Fig. \ref{fig2}). According to Fig. \ref{fig4}(c), the term $E_b$ dominates over the entire wavenumber range and destabilizes the flow, and the other terms are responsible for stabilizing the flow. This results in the destabilization of the flow due to buoyant effects in the fluid layer, which was also observed from the neutral stability curve (see Fig. \ref{fig3}(b)). 
\begin{figure}
\centering
\includegraphics[width=\textwidth]{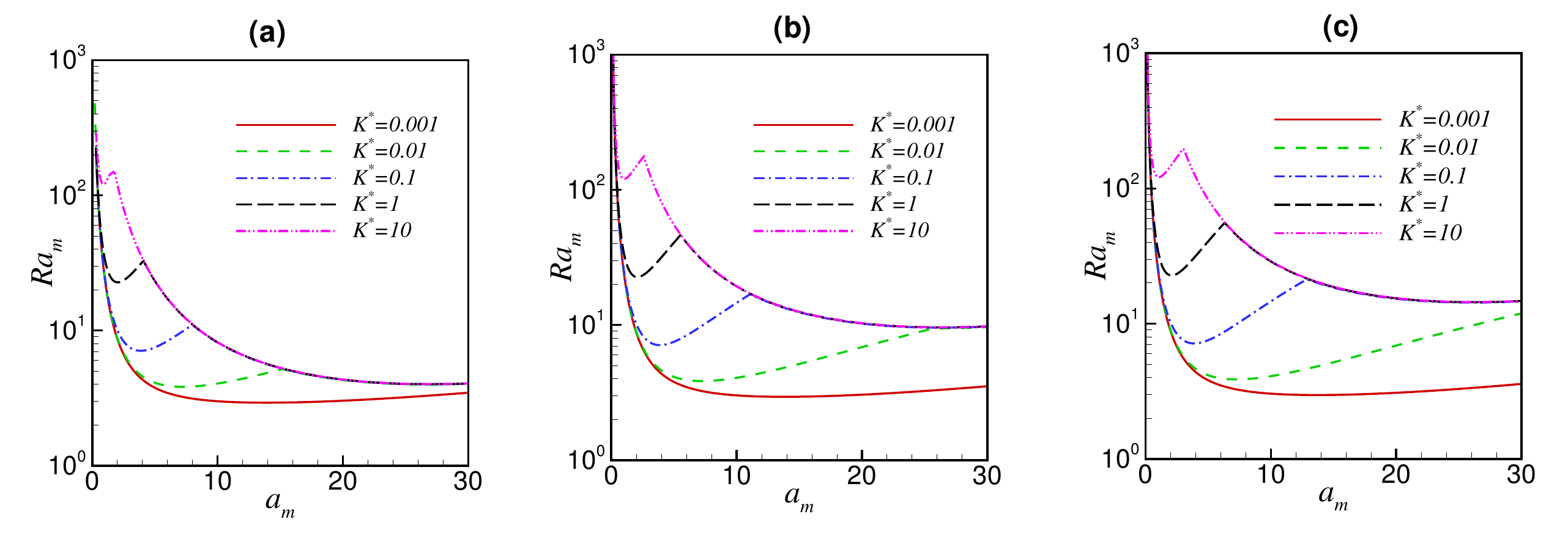}
\caption{Variation of neutral curves for different anisotropy parameter for (a) $\delta=0.0005$ (b) $\delta=0.0008$ (c) $\delta=0.001$ with $\hat{d}=0.1$, $Re=10$ and $Pr=6.9$.}
\label{fig5}
\end{figure}
\par To observe the effect of anisotropy parameter for different permeable porous medium, three different values $0.0005$, $0.0008$ and $0.001$ of $\delta$ \cite{Dee-Ana-Bas-15} are considered (see Fig. \ref{fig5}). It is to be noted here that the effect of the Darcy number finds significance in the present study even for isotropic and homogeneous porous medium since it was not yet explored till date. At $\delta=0.0005$, a unimodal neutral curve is obtained only for $K^*=0.001$, and in that case, porous mode dominates the instability. For the increasing value of $K^*$, the neutral curve becomes bimodal, and fluid mode dominates the instability for $K^*>0.01$. As $\delta$ increases to $0.0008$, the modal behavior of the neutral curve does not change, but the instability in the porous layer increases, and dominating behavior of the porous mode is obtained for $K^* \leq 0.1$. For $K^*=0.01$, instability in the fluid layer vanishes gradually for increasing the value of $\delta$, and the neutral curve becomes unimodal when $\delta=0.001$. Here, we can say that decreasing  $\delta$ increases the instability in the fluid layer as it causes hindrance for the fluid to seep into the porous layer.
\begin{figure}
\centering
\includegraphics[width=\textwidth]{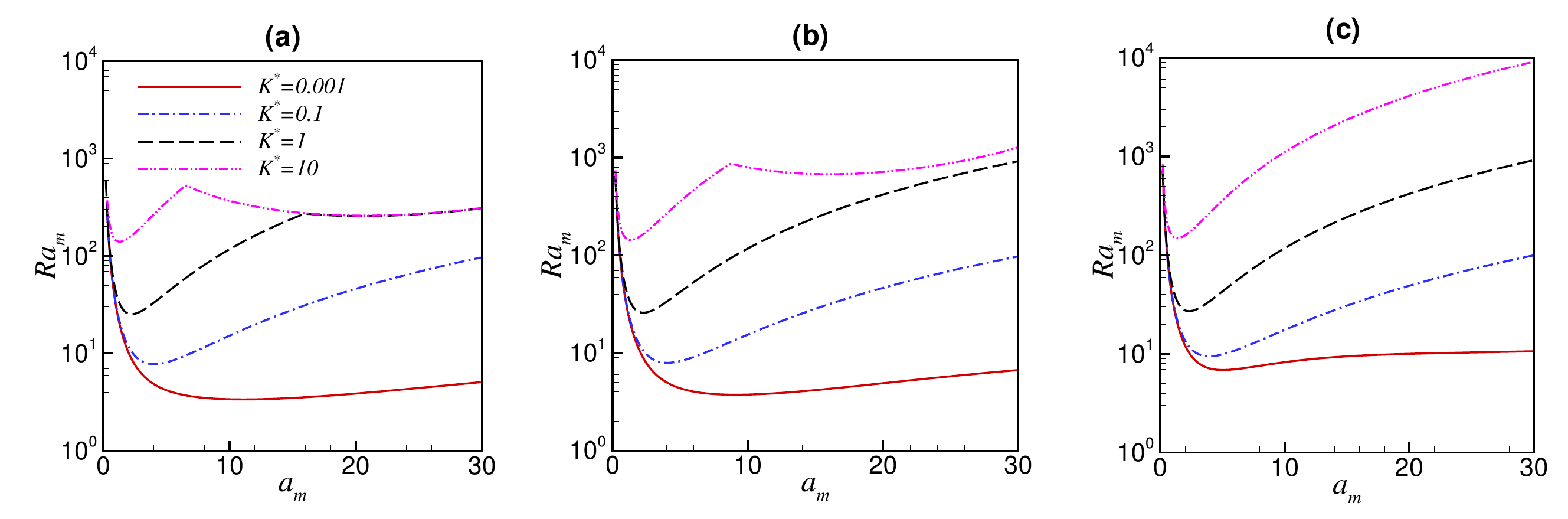}
\caption{Variation of neutral curves for different anisotropy parameter for (a) $Re=50$ (b) $Re=100$ (c) $Re=500$ with $\hat{d}=0.1$, $\delta=0.002$ and $Pr=6.9$.}
\label{fig6}
\end{figure}
\par To see the effect of shear due to the movement of the upper plate of the fluid layer for an anisotropic porous medium, three different values, $50$, $100$ and $500$ of $Re$, are considered. Here, it is noticed that the neutral curve is always unimodal with porous mode, and there is no such variation for $K^*=0.001$ and $K^*=0.01$ at different values of $Re$. So, in the remaining study, $K^*=0.01$ is omitted. As $K^*$ increases from $0.1$ to $1$, the unimodal characteristic of the neutral curve changes to bimodal for $Re=50$ and the range of $a_m$ corresponding to fluid mode is $16\le am\le30$ whereas it is $6.60\le a_m\le 30$ when $K^*=10$. Here, the porous mode dominates the system's instability, which is contrary to the previous case (Fig. \ref{fig2}), where dominating behavior of the fluid mode was obtained at $K^*=10$. At $Re=100$, the bimodal behavior of the neutral curve is obtained only when $K^*=10$ and $Re=10$. As $Re$ increases to $500$, instability in the fluid layer vanishes irrespective of the value of $K^*$. So it is clear that for an increasing value of $Re$, fluid mode stabilizes, and instability in the fluid layer gradually disappears.
\begin{figure}
\centering
\includegraphics[width=\textwidth]{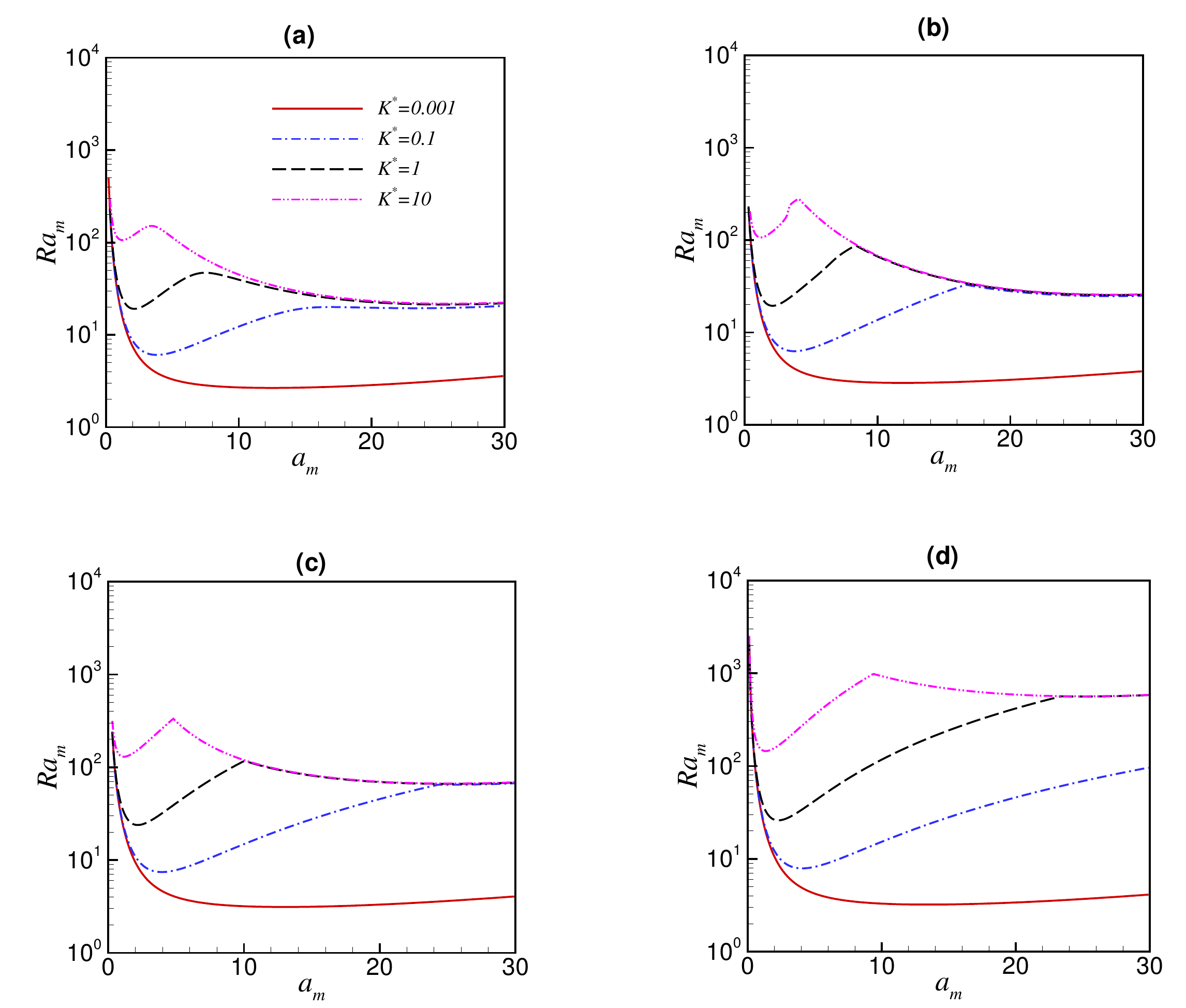}
\caption{Variation of neutral curves for different anisotropy parameter for (a) $Pr=0.01$ (b) $Pr=1$ (c) $Pr=10$ (d) $Pr=100$ with $\hat{d}=0.1$, $\delta=0.002$ and $Re=10$.}
\label{fig7}
\end{figure}
\par To investigate how anisotropy is affected for different fluids like liquid metal, air, water, and heavy oil, four values $0.01$, $1$, $10$ and $100$ of $Pr$ are considered \cite{Cha-05}. Irrespective of the value of $Pr$, the neutral curve is unimodal with porous mode when $K^*=0.001$ (shown in Fig. \ref{fig7}). For isotropic porous medium, i.e., for $K^*=1$, fluid mode comes into the picture, and as a result, the neutral curve is found to be bimodal. But still, instability is dominated by porous mode. As $K^*$ increases from $1$ to $10$, dominating behavior of fluid mode is obtained for $Pr=0.01$, $1$ and $10$. Another observation is that, for liquid metal $(Pr=0.01)$, a continuous pattern of oscillatory frequency $\sigma_m^r$ is noticed (figure not shown) when $K^*=0.1$, $1$ and $10$. In the before-mentioned cases, the range of the wavenumber corresponding to porous mode and fluid mode will be obtained from the KE balance spectrum since the exact range couldn't be deciphered from the neutral stability curves (Fig. \ref{fig7}(a)). 
\par
\begin{figure}
\centering
\includegraphics[width=\textwidth]{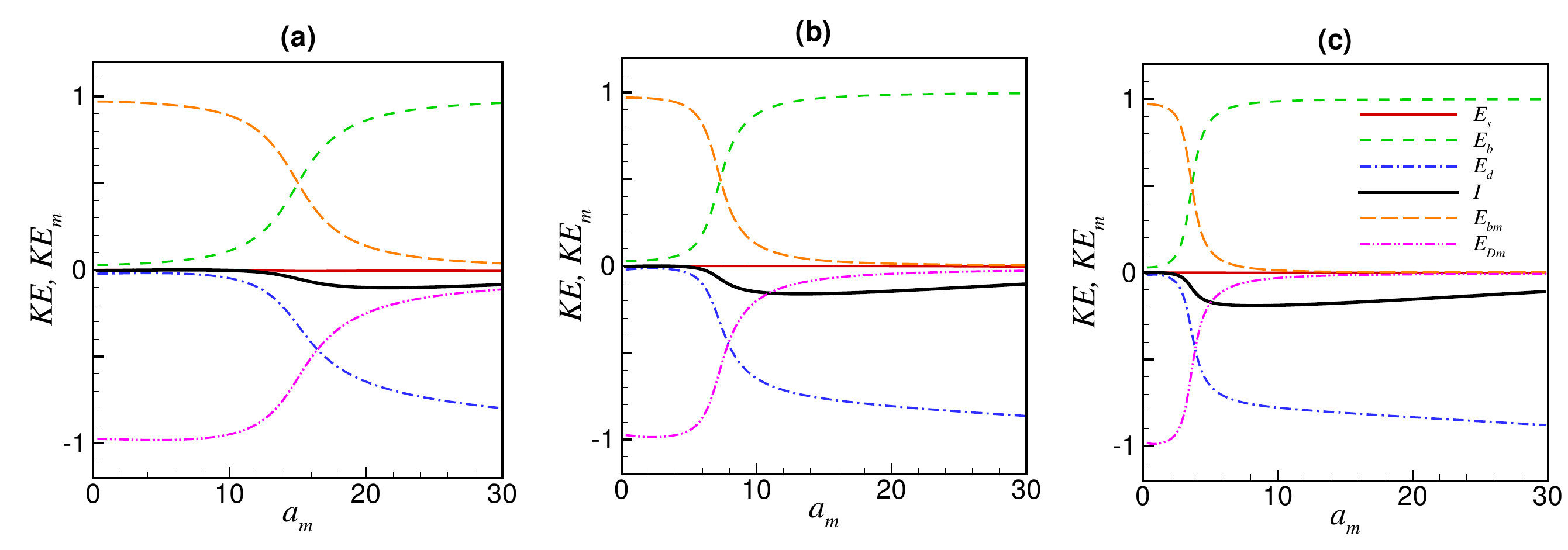}
\caption{Energy components (a) $K^*=0.1$, (b) $K^*=1$, (c) $K^*=10$ with $Pr=0.01$, $\hat{d}=0.1$, $\delta=0.002$ and $Re=10$.}
\label{fig8}
\end{figure}

 Figure \ref{fig8} represents that, in the case of liquid metal, the contribution of the energy transfer term due to buoyancy for fluid and porous layer to destabilize the flow is more than the other terms. Hence, thermal-buoyant instability occurs. Also, for $K^*=0.1$, $E_b<E_{b_m}$ in the range of wave number $0\le a_m\le 15$ so the instability is controlled by the porous layer within that range of wave number. Outside that range, the instability in the fluid layer prevails. Similarly, for $K^*=1$ and $K^*=10$, dominating characteristic of the porous layer occurs when $a_m\le 7.3$ and $a_m\le 3.6$, respectively; elsewhere, the fluid layer dominates instability. This exact range could only be calculated with the help of the observed KE spectrum.

\begin{figure}
\centering
\includegraphics[width=\textwidth]{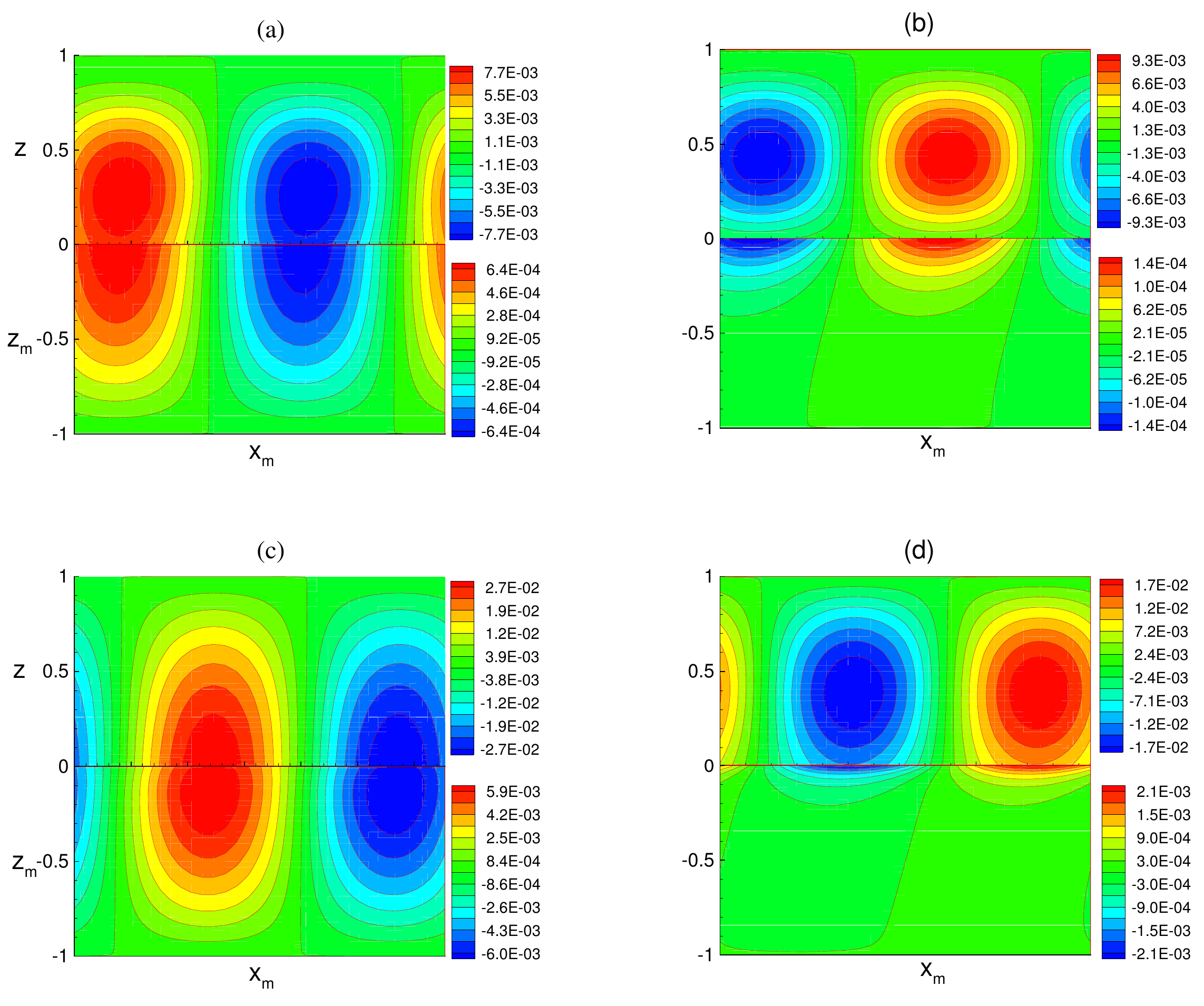}
\caption{Streamfunction pattern (a, b) along with corresponding temperature profile (c, d) for (a, c) $a_m=9.6$, $\hat{d}=0.1$, $\delta=0.002$, $Re=10$, $Pr=0.01$ and $K^*=0.1$ (b, d) $a_m=21.7$, $\hat{d}=0.1$, $\delta=0.002$, $Re=10$, $Pr=0.01$ and $K^*=0.1$.}
\label{fig9}
\end{figure}

Further, what could be the possible flow pattern when the transition of mode takes place from porous to fluid? The answer to the same is given by plotting the secondary flow patterns in terms of steamfunction and temperature contours. Figure \ref{fig9} illustrates the secondary flow pattern for different values $(0.1, 10, 0.01, 0.002, 0, 0.1)$, of the parameters $(\hat{d}, Re, Pr, \delta, A, K^*)$. The wavelength for the porous layer is scaled by $\hat{d}$ times the fluid layer on the horizontal axis. Along the vertical axis, the porous layer extends from $-1$ to $0$, and the fluid layer extends from $0$ to $1$. The system in this study is heated from below, and the temperature of the upper plate is less than that of the lower plate, so initially, heat transfer is due to conduction from bottom to top and from left to right by the movement of the upper plate. The basic flow becomes unstable when the temperature difference exceeds a threshold value, and hence convection occurs in the system. Figure \ref{fig9}(a) shows that, at $a_m=9.6$, stream function patterns are almost equally distributed in both the fluid and porous layer, and the dominating nature of porous mode is noticed, which is also verified by kinetic energy spectrum \ref{fig8}(a). Figure \ref{fig9}(b) shows that for $a_m=21.7$, the pattern of stream function is mostly within the fluid layer, which implies that fluid mode dominates instability. The corresponding temperature contour is displayed in Fig. \ref{fig9}(c,d). The temperature contours spread out evenly in the fluid and porous layer for porous mode and are majorly confined to fluid layer for fluid mode. This helps to validate the observations obtained from the preceding sections.

So far, our analysis was limited to homogeneous porous media; however, as mentioned in the literature \cite{Dee-Ana-Bas-15, anjalikhanbera2022, doomra2022effect}, the role of inhomogeneity on the interface location, media permeability, movement of the upper plate, and the type of fluid may be significant.    
\subsection{Effect of variation in inhomogeneity}
\begin{figure}
\centering
\includegraphics[width=\textwidth]{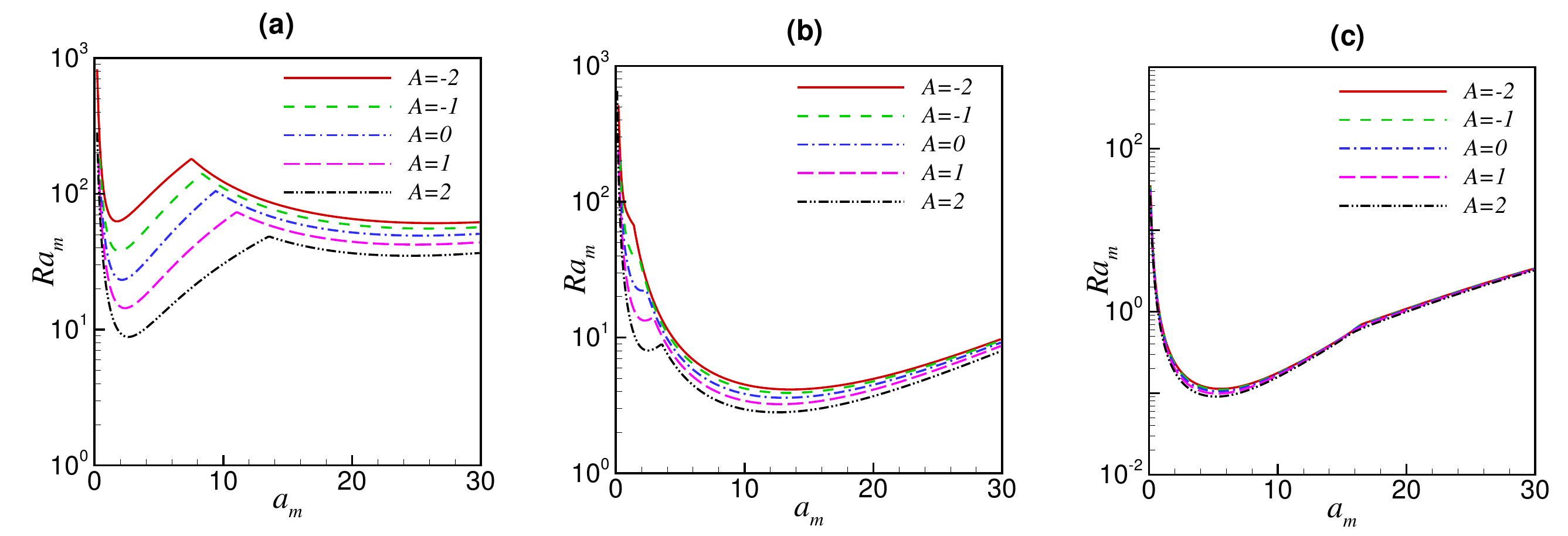}
\caption{Variation of neutral curves for different inhomogeneity parameter for (a) $\hat{d}=0.1$ (b) $\hat{d}=0.2$ (c) $\hat{d}=0.5$ with $\delta=0.002$, $Re=10$ and $Pr=6.9$.}
\label{fig10}
\end{figure}

The current section now addresses the impact of inhomogeneity variation by keeping $K^* = 1$, i.e., isotropic porous medium. Inhomogeneity parameter $A$ takes the values $-2$, $-1$, $0$, $1$, and $2$ \cite{Dee-Ana-Bas-15}. Figure \ref{fig10} represents the variation of the neutral curve for different values of inhomogeneity parameter $A$ at three different values $0.1$, $0.12$ and $0.5$ of $\hat{d}$ with $\delta=0.002$ and $Re=10$ for water. At $\hat{d}=0.1$, the neutral curve is always bimodal, and instability is dominated by porous mode except for $A=-2$. Here, it is noticed that for increasing value of $A$, the critical value of porous Rayleigh number, i.e., $Ra_{m_c}$, decreases, i.e., the porous layer becomes more unstable for increasing value of $A$. As $\hat{d}$ increases to $0.2$, the modal behavior of the neutral curve remains the same and fluid mode controls the system's instability irrespective of the value of $A$. Further increasing the value of $\hat{d}$ to $0.5$, neutral curves exhibit unimodal behavior, i.e., the fluid mode for all values of $A$. Here, it is noticed that for increasing value of $\hat{d}$, the impact of the inhomogeneity parameter becomes less significant.
\begin{figure}
\centering
\includegraphics[width=\textwidth]{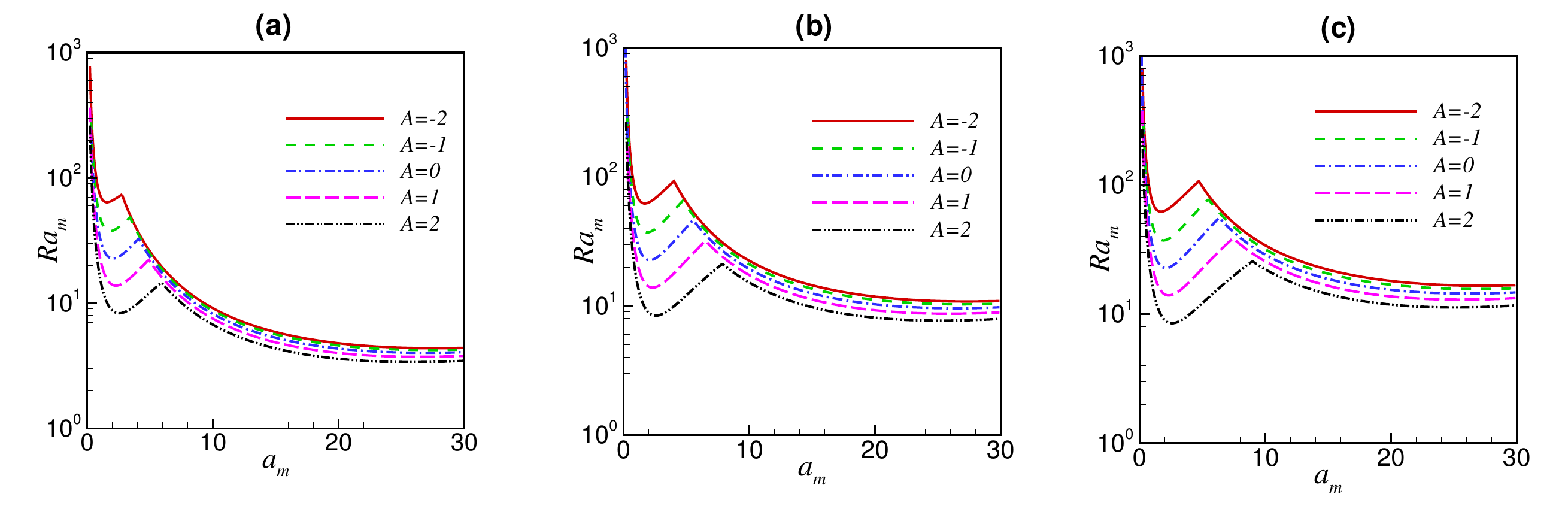}
\caption{Variations of neutral curves for different inhomogeneity parameter for (a) $\delta=0.0005$ (b) $\delta=0.0008$ (b) $\delta=0.001$ with $\hat{d}=0.1$, $Pr=6.9$ and $Re=10$.}
\label{fig11}
\end{figure}
\par
\begin{figure}
\centering
\includegraphics[width=\textwidth]{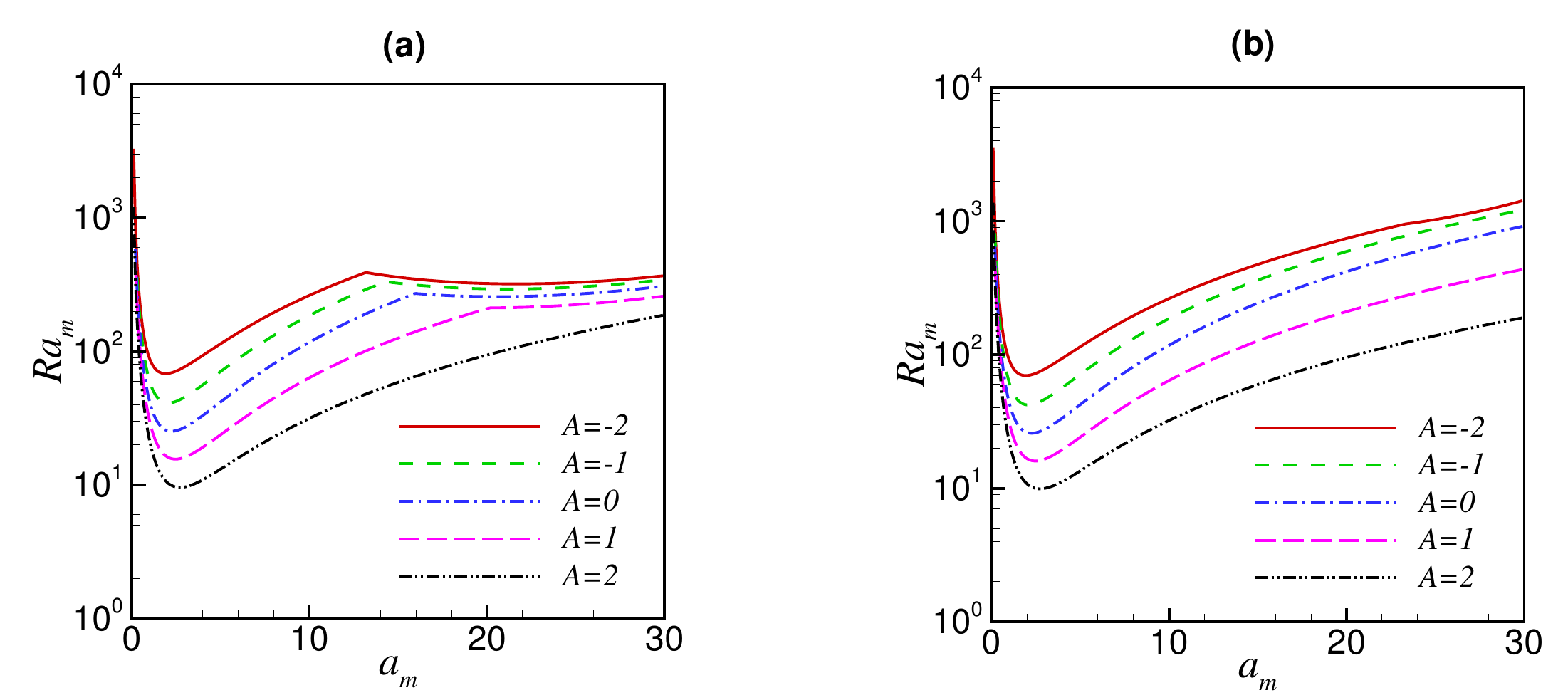}
\caption{Variation of neutral curves for different inhomogeneity parameter for (a) $Re=50$ (b) $Re=100$ with $\hat{d}=0.1$, $\delta=0.002$ and $Pr=6.9$.}
\label{fig12}
\end{figure}

 Figure \ref{fig11} represents neutral curves for several values of inhomogeneity parameter at $\delta=0.0005$, $0.0008$ and $0.001$. Here, the neutral curve is bimodal irrespective of values of $A$ as well as $\delta$. For $\delta=0.0005$ and $0.0008$, the dominant mode of instability is always fluid mode. As $\delta$ increases to $0.001$, instability of the porous layer increases, and it dominates instability only when $A=2$. In Fig. \ref{fig12}, neutral curves are shown for the different magnitudes of inhomogeneity parameter for different $Re$. At $Re=50$, it is observed that as $A$ increases, instability in the fluid layer decreases, and it vanishes when $A=2$. Regardless of the value of $A$, the porous mode dominates the instability of the system. The disappearance of instability in the fluid layer is observed for increasing $Re$ to $100$, and the porous mode appears as the only persistent mode. The trend is similar for $Re=500$ as well.

\begin{figure}
\centering
\includegraphics[width=\textwidth]{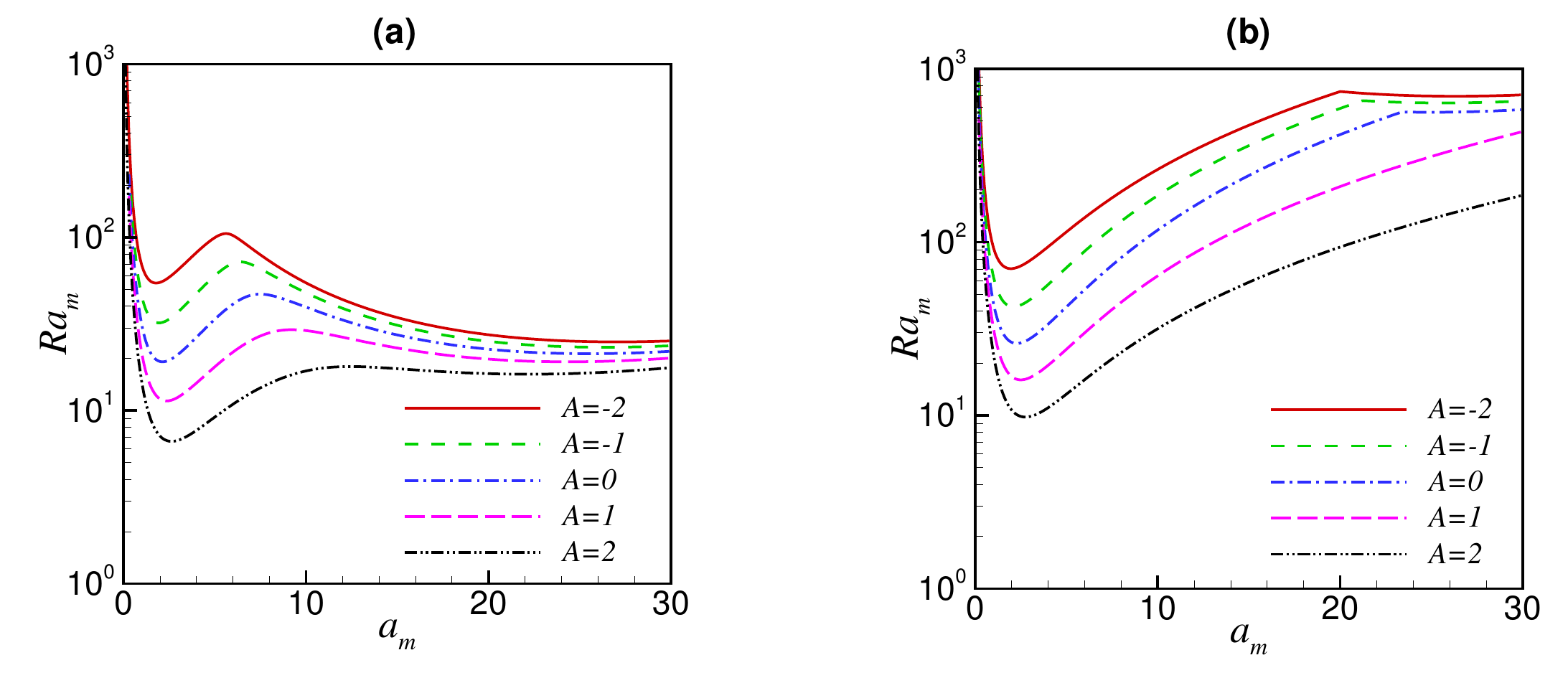}
\caption{Variation of neutral curves for different inhomogeneity parameter for (a) $Pr=0.01$ (b) $Pr=100$ with $\hat{d}=0.1$, $\delta=0.002$ and $Re=10$.}
\label{fig13}
\end{figure}

The inhomogeneity effect on different fluid types is shown in Fig. \ref{fig13}. For liquid metal, i.e., $Pr=0.01$, the neutral curves are observed to be bimodal and in contrast to other fluids (air, water, heavy oil); here, the oscillatory frequency varies smoothly as a function of wavenumber. It has also been observed that fluid mode controls system instability when $A=-2$ and $A=-1$. For $Pr=1$ and $10$, neutral curves exhibit bimodal behavior, and the porous mode is observed to dominate except where $Pr=1$ with $A=-2$ and $-1$ (figures not shown). As $Pr$ increases, instability in the porous layer increases. Unless $A=-2,-1,0$, the neutral curve is unimodal at $Pr=100$. Moreover, the instability is primarily caused by porous mode regardless of values of $A$.
\subsection{Combined effect of variation in anisotropy and inhomogeneity}
\begin{figure}
\centering
\includegraphics[width=\textwidth]{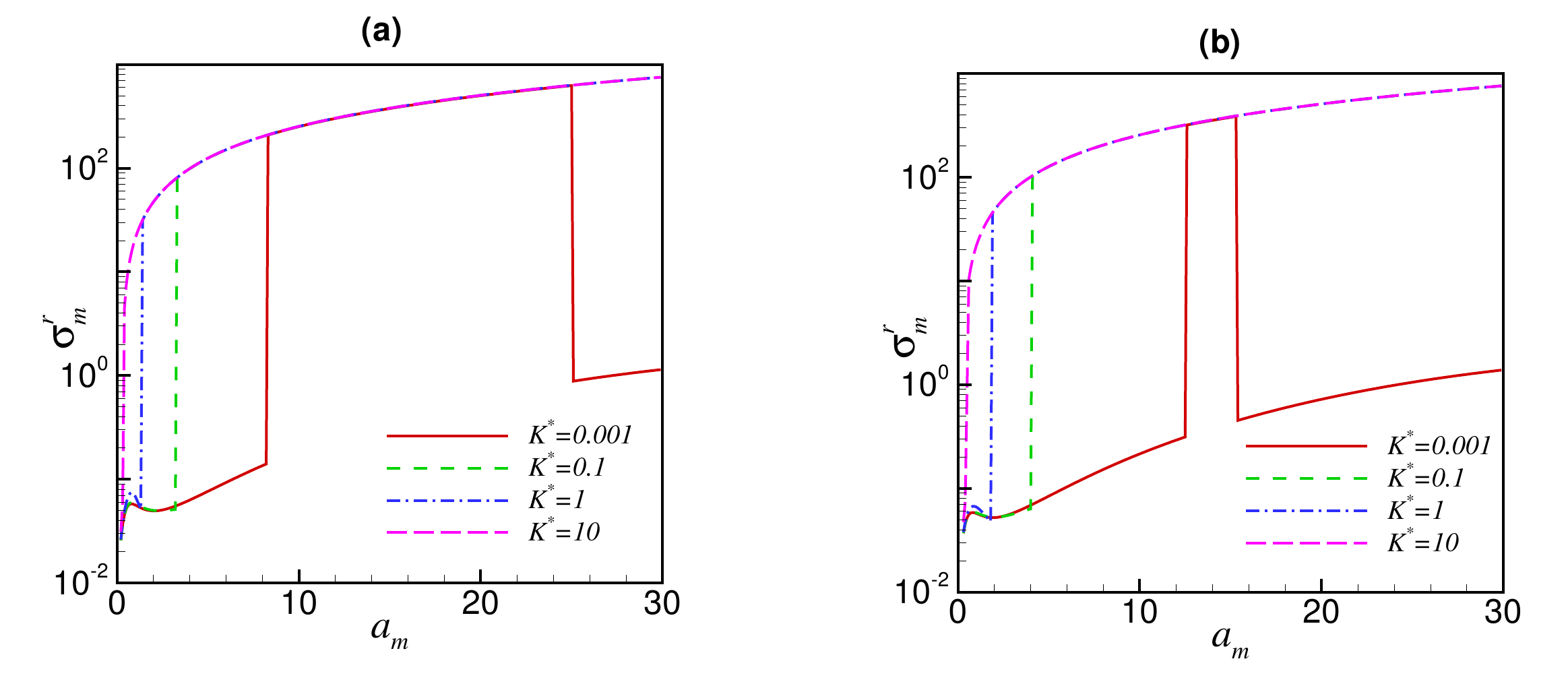}
\caption{Variation of neutral curves for different anisotropy parameter for (a) $A=-2$ (b) $A=-1$ with $\hat{d}=0.2$, $\delta=0.002$, $Pr=6.9$ and $Re=10$.}
\label{fig14}
\end{figure}

\begin{figure}
\centering
\includegraphics[width=\textwidth]{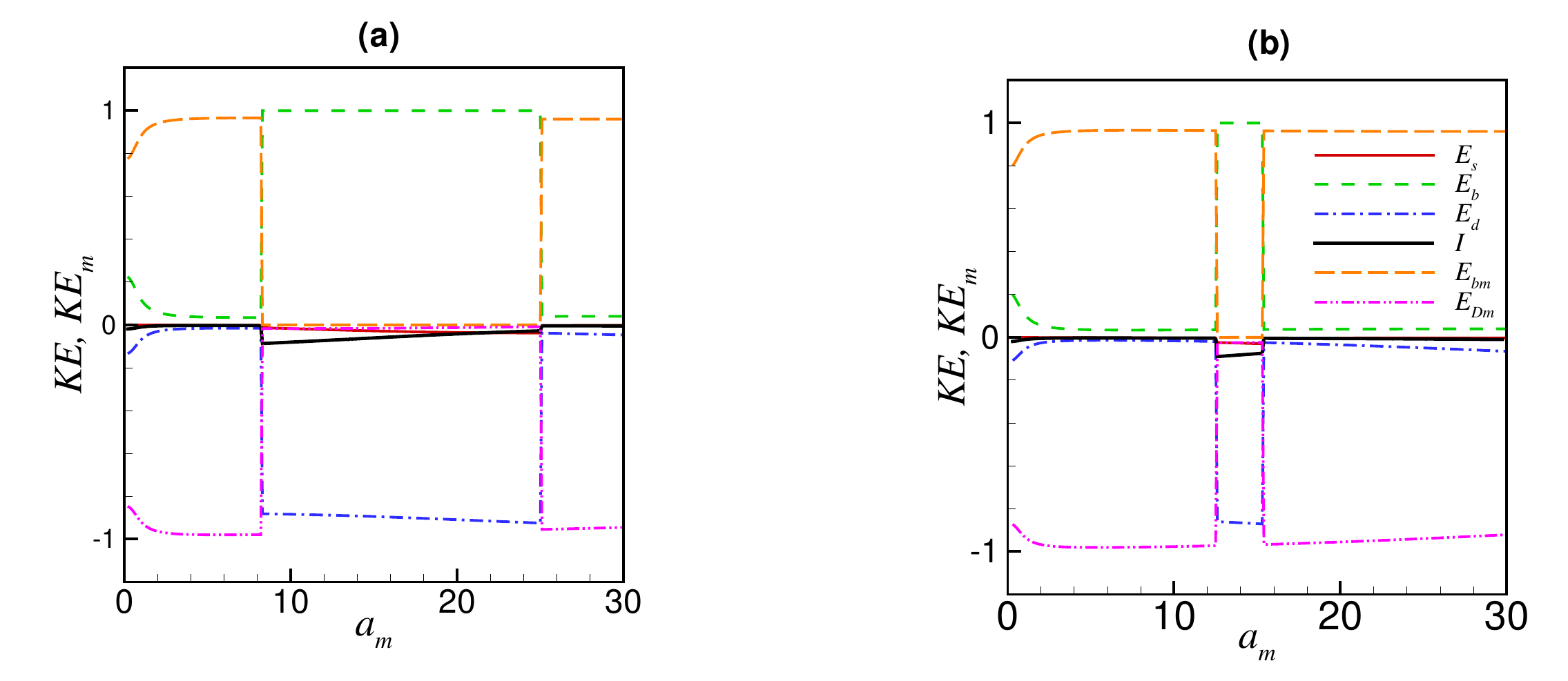}
\caption{Energy components for (a) $A=-2$ (b) $A=-1$ with $\hat{d}=0.2$, $\delta=0.002$, $Re=10$ and $Pr=6.9$.}
\label{fig15}
\end{figure}

\begin{figure}
\centering
\includegraphics[width=\textwidth]{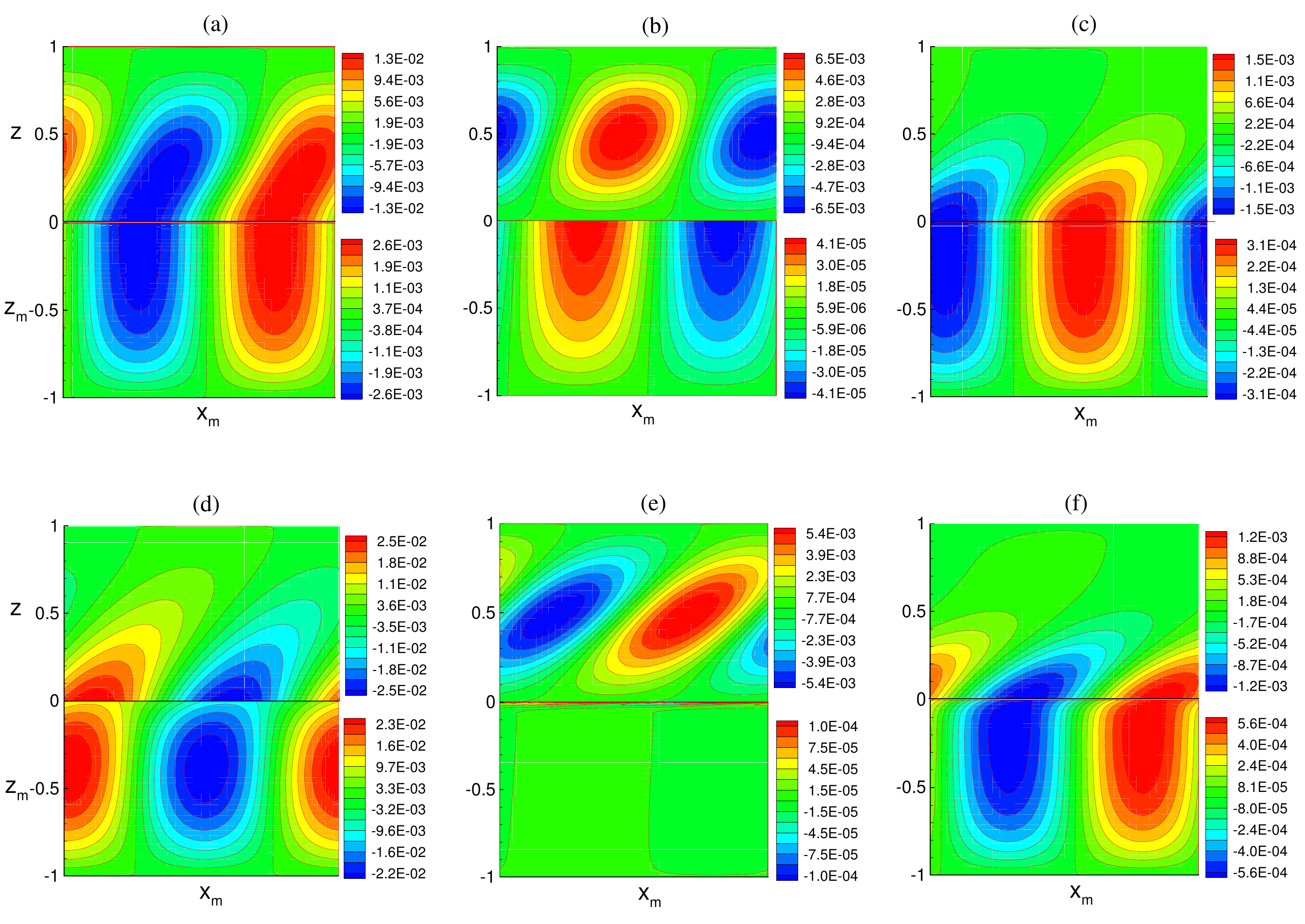}
\caption{Streamfunction pattern (a, b, c) along with corresponding temperature profile (d, e, f) for (a, d) $a_m=3.7$, $\hat{d}=0.2$, $\delta=0.002$, $Pr=6.9$, $Re=10$, $K^*=0.001$ and $A=-2$ (b, e) $a_m=20.5$, $\hat{d}=0.2$, $\delta=0.002$, $Pr=6.9$, $Re=10$, $K^*=0.001$ and $A=-2$ (c, f) $a_m=27.10$, $\hat{d}=0.2$, $\delta=0.002$, $Pr=6.9$, $Re=10$, $K^*=0.001$ and $A=-2$.}
\label{fig16}
\end{figure}

The previous sections dealt with the effect of anisotropy and inhomogeneity independently. However, to understand the simultaneous impact of the above parameters on the instability mechanism, Figs. \ref{fig14}-\ref{fig16} are analyzed. For the fluid as water, the change in the neutral curve for various values of $K^*$ is shown in Fig. \ref{fig14}(a) and \ref{fig14}(b) at $A=-2$ and $-1$, respectively with $\hat{d}=0.2$, $\delta=0.002$ and $Re=10$. Interestingly, the simultaneous variation of anisotropy and inhomogeneity introduces trimodal instability. In trimodal characteristic, the least stable mode shifts from porous to fluid and then from fluid to porous as the wavenumber increases. At $A=-2$, porous mode dominates instability for $0<a_m\le 8.2$. Thereafter, the dominating nature of fluid mode is observed for $8.2<a_m\le 25$, and further enhancement of $a_m$ introduces porous mode again for $a_m>25.$ Similarly, at $A=-1$, the first mode is the porous mode for $a_m\le 12.50$, then fluid mode for $12.50<a_m\le 15.30$ and then again porous mode for $a_m>15.30$. We have also plotted the respective kinetic energy spectrum in Fig. \ref{fig15} to understand the underlying physical mechanism of the trimodal instability. Here, it is seen that, only the energy transfer term owing to buoyancy in fluid and porous layer is positive throughout the range of wave number, which acts as a destabilizing factor. From Fig. \ref{fig14}, the thermal buoyant instability in the porous layer is observed when $E_{bm}>E_b$ with $a_m\le 8.2$ and $a_m> 25$. In the other range, $E_b$ dominates instability, and that indicates thermal buoyant instability in the fluid layer. The negative shear indicates the loss of the disturbance kinetic energy to the mean flow through Reynolds shear stress, referred to as the shear destruction \cite{ber-kha-2006}. The destabilizing KE is balanced mainly by surface drag $(E_{Dm})$in porous mode, whereas, for fluid mode, it is balanced by the dissipation of KE in the fluid layer $(E_d)$. Furthermore, to understand the variation of the pattern of secondary flow as the mode of instability changes from porous to fluid and then from fluid to porous, we have plotted the same at chosen respective wavenumbers $3.7, 20.5$ \& $27.10$ for $A=-2$ in Fig. \ref{fig16}. Figure \ref{fig16}(a)-(c) shows that the stream function contours are spread over both fluid and porous regions. Figure \ref{fig16}(a) and \ref{fig16}(c) show relatively high flow convection in the porous layer indicating the dominance of porous mode in the system, whereas the convection is dominant in the fluid region in Fig. \ref{fig16}(b), which signifies fluid mode in this case. The corresponding temperature contours are shown in Fig. \ref{fig16}(d)-(f). It can be seen that the temperature contours corresponding to the fluid mode occupy the entire fluid layer and form complete elliptical circuits in the fluid layer indicating its higher strength in the respective layer. The same goes for the chosen wavenumber from porous mode, where the strength of the flow via the temperature contours finds its dominance in the porous layer. Apart from these, the strength of secondary flow (in terms of magnitude) for the same mode of instability is relatively low as the wavenumber is increased. These flow patterns also provide a validation to our results from the linear stability and energy spectrum.

\par In contrast to non-isothermal Poiseuille flow in superposed system  \cite{anjalikhanbera2022}, where bimodal nature of neutral curve is obtained for small value of anisotropy parameter $(K^*=0.001)$ and inhomogeneity parameter $(A=-1)$ and $\hat{d}=0.13,$ $Pr=10,$ $Re=10$ and $\delta=0.001$, here, in the Couette flow unimodal (porous mode) nature of neutral curve is observed. For $A=0$ and $1$, the instability of the plane Poiseuille flow is controlled by fluid mode when $\hat{d}=0.1,$ $Pr=100,$ $Re=10$, $\delta=0.001$ and $K^*=10$ whereas, for the same set of parametric values, the instability of Couette flow is dominated by porous mode. 

\section{Conclusions}\label{sec4}
The linear stability of non-isothermal plane Couette flow in the fluid overlying anisotropic and inhomogeneous porous layer using the two-domain approach is analyzed in the present study. To discretize the linearized perturbed equations, the Chebyshev collocation method is used, and by adopting the QZ algorithm; the generalized eigenvalue problem is solved. The present study has been validated with the theoretical and experimental results in the limit of the isotropic and homogeneous porous layer with the results of \cite{Cha-05} and \cite{Che-Che-89} ($Re \to 0$), respectively. The effect of depth ratio, anisotropy, inhomogeneity, Reynolds number, Darcy number and Prandtl number are discussed in this article.  

It is found that the media anisotropy and inhomogeneity do not alter the stability characteristic of the isothermal plane Couette flow in the fluid overlying porous layer, i.e., it still remains unconditionally stable. On viewing the influence of anisotropy and inhomogeneity for non-isothermal Couette flow, the change from unimodal (porous mode) to bimodal and back to unimodal (fluid mode) is noticed as $\hat{d}$ increases. For the higher value of $\hat{d}$, the effect of inhomogeneity and anisotropy becomes insignificant, and instability in the porous layer dies out for the considered range of parameters. It has also been observed that when the underlying porous media is saturated with water, for $\delta=0.002$ and $Re=10$, porous mode always dominates instability for $\hat{d}<0.07$, whereas, fluid mode controls the instability for $\hat{d}>0.21$ irrespective of $A$ and $K^*$. It is seen that porous mode becomes more unstable for decreasing value in $K^*$ and bimodal neutral curves shift to unimodal (porous mode). The critical Rayleigh porous number increases with the increasing value of the inhomogeneity parameter, which indicates that the system becomes more unstable with increasing the inhomogeneity parameter. As $\delta$, $Re$ and $Pr$ increase, the bimodal nature of the neutral curve changes to unimodal and gradually, instability in the fluid layer vanishes. 

Contrary to the unimodal and bimodal nature of the neutral curve for homogeneous and isotropic porous medium \cite{Cha-05}, trimodal instability is obtained for the values $(0.2, 10, 6.9, 0.002, 0.001, -2)$ and $(0.2, 10, 6.9, 0.002,$ $0.001, -1)$ of the parameters $(\hat{d}, Re, Pr, \delta, K^*, A)$ in the present study. Except for liquid metal, the mode of instability changes suddenly for the considered parametric variation. The KE spectrum reveals that in balancing the destabilizing kinetic energy, energy due to surface drag (dissipation) acts as a main stabilizing factor for porous mode (fluid mode). From the KE spectrum, it is obtained that thermal buoyant instability in fluid and porous layer are the dominant factors that play a key role in driving the instability in the system. Also, time and again, secondary flow patterns are visualized to understand the flow dynamics.

The present study is focused on linear stability analysis and to go beyond the scope of this analysis, a weakly non-linear stability analysis is required. Further, the non-linear stability analysis can help to understand the transition to turbulence more clearly. Along these lines, the non-linear stability analysis of the present study is in progress.

\section*{Acknowledgement}
The financial support is provided by the Ministry of Human Resources and Development (MHRD), India and SERB, India (project grant no. EEQ/2020/000101).
\appendix

\section{Variation of neutral curve for different $K^*$ taking $J=0$ and $J=1$}\label{sec6}
\begin{figure}
\centering
\includegraphics[width=0.5\textwidth]{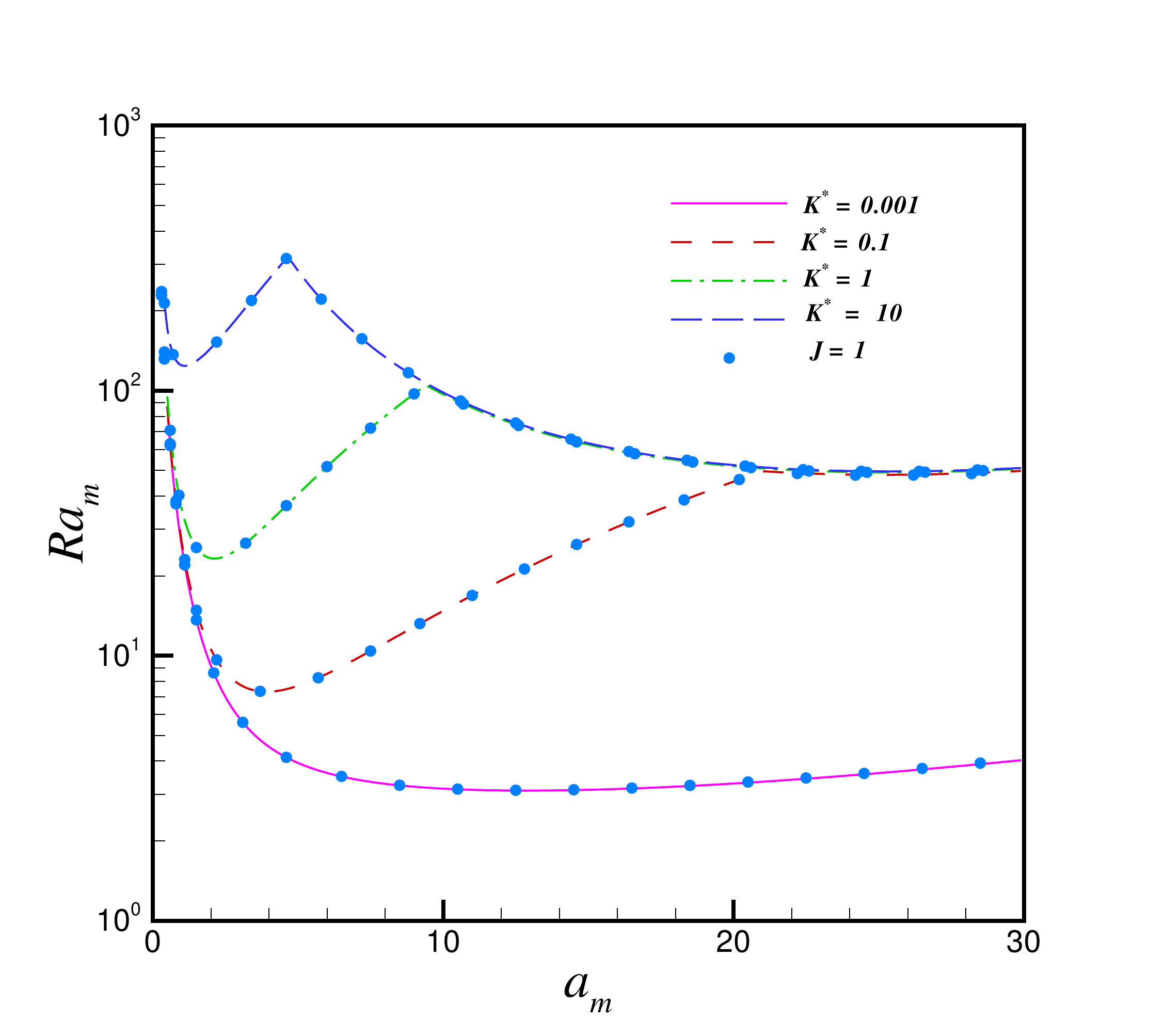}
\caption{Variation of neutral curve for different $K^*$ for $J=0$ (lines) and $J=1$ (dotted) with $\hat{d}=0.1$, $\delta=0.002$, $Re=10$, $A=0$ and $Pr=6.9$.}
\label{fig18}
\end{figure}

Figure \ref{fig18} represents the variation of neutral stability curve for different values of $K^*$ by considering the Jones condition $(J=1)$ and Beavers-Joseph condition $(J=0)$ at the fluid-porous interface.

\section{Variation of neutral curve for different $A$ and $B$}\label{sec5}
\begin{figure}
\centering
\includegraphics[width=\textwidth]{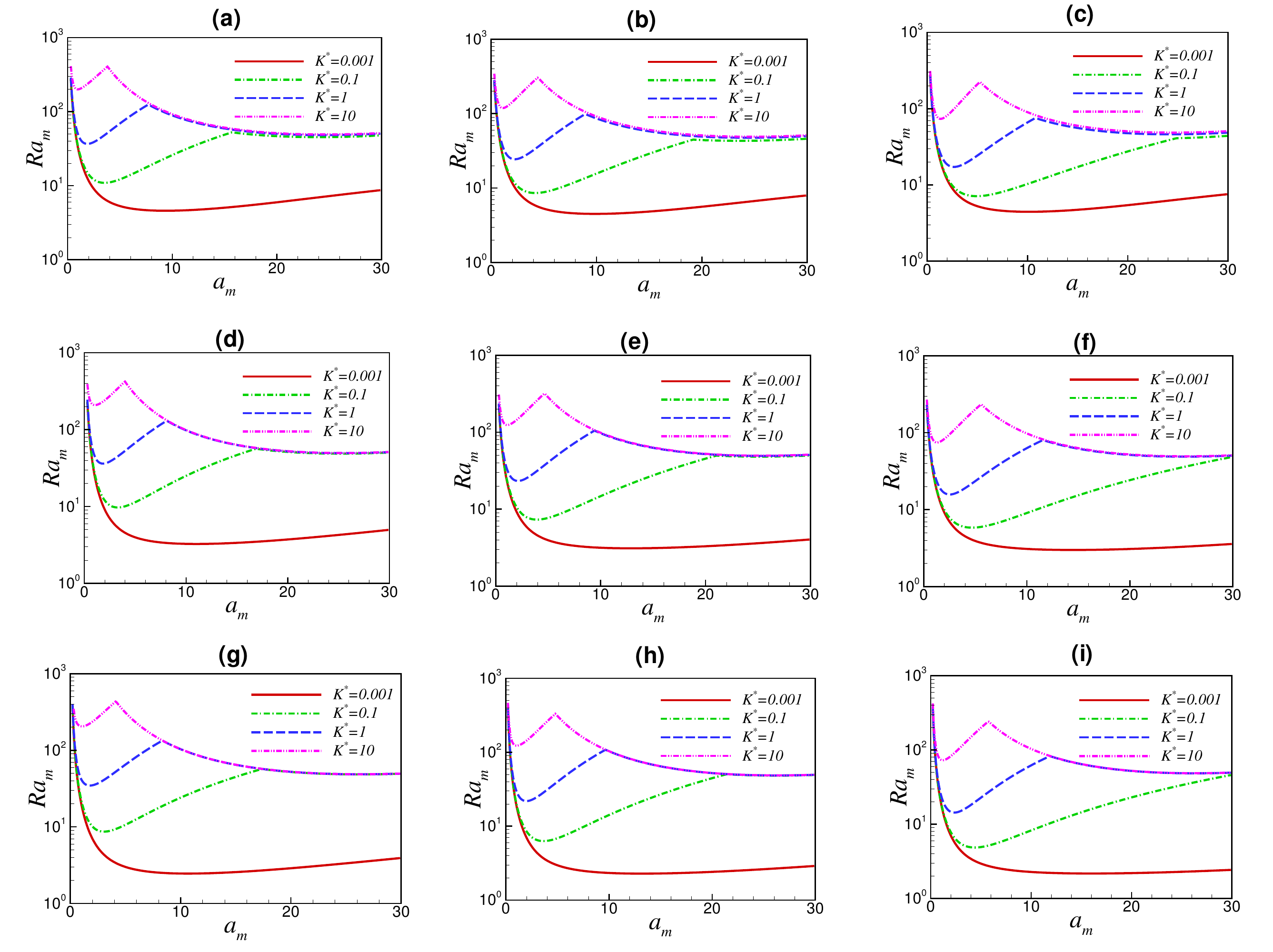}
\caption{Variation of neutral curve for (a) $A=-1$, $B=-1$ (b)  $A=-1$, $B=0$ (c)  $A=-1$, $B=1$ (d) $A=0$, $B=-1$ (e)  $A=0$, $B=0$ (f)  $A=0$, $B=1$ (g) $A=1$, $B=-1$ (h)  $A=1$, $B=0$ (i)  $A=1$, $B=1$ with $\hat{d}=0.1$, $\delta=0.002$, $Re=10$ and $Pr=6.9$.}
\label{fig17}
\end{figure}
Considering different inhomogeneity function $\eta_x = e^{A(1+z_m)}$ and $\eta_z = e^{B(1+z_m)}$ along horizontal and vertical direction respectively, variation of the neutral stability curve for different values of anisotropy parameter $(0.001, 0.1, 1, 10)$ and inhomogeneity parameter $(-1, 0, 1)$ are shown in Fig. \ref{fig17}.

\end{document}